\documentclass[epj]{svjour}
\usepackage{graphicx}
\usepackage{dcolumn}
\usepackage{bm}
\usepackage{amsmath,mathtools,amssymb}

\newcommand{\pt}{$p_\mathrm{T}$}
\begin{document}
\title{A realistic coalescence model for deuteron production}
\author{Maximilian Mahlein\inst{1,6} \and Luca Barioglio\inst{2,} \inst{1} \and Francesca Bellini\inst{3,4} \and Laura Fabbietti\inst{1} \and Chiara Pinto\inst{1} \and Bhawani Singh\inst{1} \and Sushanta Tripathy\inst{4,5}
}                     % Do not remove
\institute{Technical University of Munich, TUM School of Natural Sciences, Physics Department, James-Franck-Stra{\ss}e 1, 85748 Garching b. M\"{u}nchen, Germany \and INFN, Sezione di Torino, Via Pietro Giuria 1, Torino, 10125, TO, Italy \and University of Bologna, Department of Physics and Astronomy A. Righi, Via Irnerio 46, Bologna, 40126, BO, Italy \and INFN, Sezione di Bologna, Viale Berti Pichat 6/2, Bologna, 40127, BO, Italy \and CERN, 1211 Geneva, Switzerland \and maximilian.horst@tum.de }
\date{Received: date / Revised version: date}
\abstract{
A microscopic understanding of (anti)deuteron production in hadron-hadron collisions is the subject of many experimental and theoretical efforts in nuclear physics. This topic is also very relevant for astrophysics, since the rare production of antinuclei in our Universe could be a doorway to discover new physics.
In this work, we describe a new coalescence afterburner for event generators based on the Wigner function formalism and we apply it to the (anti)deuteron case, taking into account a realistic particle emitting source. The model performance is validated using the EPOS and PYTHIA event generators applied to proton-proton collisions at the centre-of-mass energy $\sqrt{s}=$ 13 TeV, triggered for high multiplicity events, and the experimental data measured by ALICE in the same collision system. The model relies on the direct measurement of the particle emitting source carried out by means of nucleon-nucleon femtoscopic correlations in the same collision system and energy. The resulting model is used to predict deuteron differential spectra assuming different deuteron wavefunctions within the Wigner function formalism. The predicted deuteron spectra show a clear sensitivity to the choice of the deuteron wavefunction. The Argonne $v_{18}$ wavefunction provides the best description of the experimental data. This model can now be used to study the production of (anti)deuterons over a wide range of collision energies and be extended to heavier nuclei.
%\PACS{
%      {PACS-key}{discribing text of that key}   \and
 %     {PACS-key}{discribing text of that key}
%     } % end of PACS codes
} %end of abstract
\maketitle
\section{Introduction}
\label{sec:introduction}
The microscopic understanding of the formation of light nuclei in high-energy collisions is a fundamental open problem that is being addressed since several decades both experimentally and theoretically.
The main question is how nuclear bound states are formed and their structure emerges from the properties of the strong interaction and the laws of Quantum Chromodynamics. This problem is also relevant in astroparticle physics, where insights on the production mechanism are necessary to interpret future measurements of antinuclei in cosmic rays by space-borne experiments searching for dark matter~\cite{Korsmeier:2017xzj,Blum:2017qnn,Kachelriess:2020uoh,vonDoetinchem:2020vbj,PhysRevD.105.083021}. 
Indeed, it has been suggested that, among the products of dark matter particle annihilation and decay, light antinuclei could be a promising signal to search for, since the only known background source are cosmic rays interactions with the interstellar medium.
As cosmic rays and the interstellar medium mostly consist of hydrogen (90$\%$) and helium (8$\%$), and only in small percentage of heavier nuclei, most of the relevant interactions for the production of antinuclei in the Galaxy are proton-proton (pp) and proton–helium collisions. Studying these processes in the laboratory provides a unique opportunity to emulate under controlled conditions the production of antinuclei in the Galaxy. 

The production of light nuclei has been measured at fixed-target and collider experiments using pp and nuclear collisions up to Pb--Pb over a broad range of beam energies, ranging from the AGS~\cite{Bennett:1998be,Ahle:1999in,Armstrong:2000gz,Armstrong:2000gd}, to the SPS~\cite{Ambrosini:1997bf}, RHIC~\cite{Adler:2001prl,Adler:2004uy,Arsene:2010px,Agakishiev:2011ib,Adamczyk:2016gfs,Adam:2019wnb}, and the LHC~\cite{nuclei_pp_PbPb,nuclei_pp,deuteron_pp_7TeV,deuteron_pPbALICE,3He_pPb,deuteron_pp_13TeV,nuclei_pp_5TeV,nuclei_pp_13TeV_HM}, where the maximum centre-of-mass energy of \mbox{$\sqrt{s} = 13$ TeV} is reached.

From the theory side, the production of light (anti)nuclei is mainly described by the statistical hadronisation or the coalescence approach. 
Statistical hadronisation models (SHMs) assume that nuclei are emitted from a source at hadron-chemical and thermal equilibrium and their abundances are fixed at the so-called chemical freeze-out~\cite{SHM1,SHM2,SHM3,SHM4,SHM5,SHM6}. Yields depend on the hadron mass, on the temperature and the volume of the system at freeze-out, and on the conservation of quantum numbers, which is imposed to hold on average if the volume of the produced system is large (e.g. in Pb--Pb collisions), or exactly if the volume is small (e.g. in pp collisions). The SHM is a macroscopic model: it predicts hadron yields requiring the equilibrium conditions, without any control on the microscopic production mechanism. 
Also, no detailed information on the structure of bound objects is contained in the model~\cite{Muller:2022uuv} and the predicted yields are integrated over transverse momentum. 

On the contrary, coalescence aims at a microscopic description of the formation of (anti)nuclei, being the result of final state interactions between (anti)protons and (anti)neutrons. 
Coalescence was first introduced in the 1960s to explain the production of deuterons in proton-nucleus collisions at the CERN SPS~\cite{Butler:1963,Kapusta:1980} and has been employed since then to describe the production of light nuclei in relativistic heavy--ion as well as pp collisions~\cite{Sato:1981ez,Nagle:1996vp,Scheibl:1998tk,Blum:2017qnn,Blum:2019suo,Mrowczynski:2019yrr,Bellini:2020cbj}. In this approach, nucleons that are close to each other in space and momentum can coalesce to form a nucleus~\cite{Butler:1963,Kapusta:1980} when the system produced in high-energy hadronic or nuclear interactions undergoes freeze-out. 
The early works of Pearson and Butler~\cite{Butler:1963}, and Kapusta~\cite{Kapusta:1980} emphasized the momentum dependence of the coalescence probability, by identifying the coalescence momentum $p_0$ as the main parameter governing the probability of cluster formation. 
Sato and Yazaki~\cite{Sato:1981ez} introduced the density matrix formalism to calculate the nucleus formation probability, given by the integral over coordinate space of the projection of the proton and neutron wavefunctions over the deuteron one. They assumed no correlation between spatial and momentum distributions, nor any dynamical correlation between the proton and the neutron emitted from the particle source created after the hadron-hadron collision. 

Following works~\cite{Nagle:1996vp,Scheibl:1998tk} drew attention to the constraint in coordinate space to be considered in heavy-ion collisions, where the source is a spatially-extended and collectively-expanding system. In their seminal work, Scheibl and Heinz~\cite{Scheibl:1998tk} included a detailed modelling of the system and pointed out that the source could be identified with the volume out of which particles are emitted with similar momenta, the size of which can be extracted by measuring two-particle momentum correlations with femtoscopy techniques~\cite{Lednicky:1981su,Heinz:1999rw,Lisa:2005dd}. 
In recent years, a similar approach was employed to explain the production of (anti)nuclei in pp and p--Pb collisions~\cite{Blum:2017qnn,Blum:2019suo}, where the final-state particles are expected to occupy a small volume, of the order of the size of the proton ($r_p \approx 1$ fm) and even smaller than the typical size of a light nucleus (for deuteron, $r_d \approx$ 2 fm).
A key aspect of these works is having identified that the coalescence process depends on the size of the hadron emission region and more specifically, on the size of the nucleus relative to the size of the source. This relation provides the relevant length scale for the process~\cite{Blum:2017qnn,Bellini:2018epz,Bellini:2020cbj}. 

Using a Wigner function representation of nucleons and nuclei, Blum et al.~\cite{Blum:2019suo,Bellini:2020cbj} and Mrowczynski et al.~\cite{Mrowczynski:2019yrr} obtained a relation between femtoscopic correlations and coalescence, being the continuum and discrete result of final state interactions among nucleons, respectively.
Notably, they employed a Gaussian wavefunction for the \\deuteron, %this is a manual fix so it wouldnt overlap with the next column
which allows for a fully analytic calculation of the coalescence parameter.
All the mentioned models are based on the Wigner formalism and differ by the details of their implementation, including how the source size is accounted for. 
However, they have in common that they provide an analytical or numerical solution for the coalescence probability to be directly compared with experimental data. 

Motivated by the need to model the production of light antinuclei in cosmic ray interactions for the estimate of their flux near Earth, recent developments of the coalescence model focus on pp collisions, where small particle sources are expected. 
In particular, the work by Kachelriess et al.~\cite{alternativeCoalescence,KachelriessLast} provides a solution to apply coalescence as an afterburner to particle production. In their approach, Monte Carlo (MC) event generators are used to obtain two-nucleon momentum correlations from the simulated pp collision events. To form a deuteron, the Wigner function-based coalescence is employed on an event-by-event basis in the WiFunC framework~\cite{KachelriessLast}. The size of the formation region, a free parameter of the model, is taken as collision process-dependent. This parameter is fitted to ALICE data in~\cite{alternativeCoalescence}, whereas it is predicted in~\cite{KachelriessLast} within WiFunC from two-particle momentum correlations, thus being limited to the accuracy of the description of the two-nucleon correlations native to the generators. It is shown that the predicted size parameter is consistent with ALICE femtoscopy data for the baryon emitting source~\cite{sourceSizeHMpp} and is close to 1 fm, from e$^{+}$e$^{-}$ to pp collisions. 

In our work, inspired by~\cite{KachelriessLast}, we provide a coalescence afterburner that takes into account realistic particle emission and correlation, to be used to simulate event-by-event deuteron and antideuteron \footnote{We assume that the formation mechanism of antinuclei and nuclei in high-energy collisions is the same. This is justified by the present evidence that the force responsible for the nuclear binding, a residual of the strong force among partons, acts in the same way among antinucleons inside antinuclei as it does among nucleons inside matter nuclei~\cite{ALICE:2015rey}. In the following, we omit the prefix ``anti'' for brevity.} production in high-energy hadronic collisions.
We focus on pp collisions due to their relevance for the searches for antinuclei in cosmic rays as potential signatures for dark matter particles. We chose to test and validate our model by simulating deuteron production in high multiplicity pp collisions at \mbox{$\sqrt{s} = 13$~TeV} because this corresponds to the only data sample for which simultaneous measurements of deuteron yields and baryon source radius are available~\cite{sourceSizeHMpp,nuclei_pp_13TeV_HM}. As it will be evident later, the size of the emitting source is a key ingredient of our implementation, allowing us to provide a realistic model for deuteron production.

First, we generate pp collision events using two distinct MC generators, EPOS 3~\cite{Werner:2010aa,Werner:2013tya} and PYTHIA 8.3~\cite{Sjostrand:2014zea}, chosen for their known capability to reproduce most features of the LHC pp data. Then, the momentum distributions of the final state nucleons are tuned to data and are used as inputs for the afterburner. The latter is implemented based on a state-of-the-art Wigner-function coalescence approach, described in Sec.~\ref{sec:wigner}. Compared to previous approaches, we improve the generator-borne description of the nucleon-emitting source, as described in Sec.~\ref{sec:source}, by including a) a realistic account of its size derived from the measurement of two-baryon correlations, and b) a modelling of the cocktail of short-lived hadron resonances, which lead to a delayed nucleon emission. Using this improved framework, we predict the differential deuteron spectra with the Hulthén~\cite{Heinz:1999rw}, Argonne $v_{18}$~\cite{PhysRevC.51.38} and Chiral Effective Field Theory (N$^4$LO)~\cite{chiEFT} wavefunctions, moving past the traditional Gaussian wavefunction approach, and compare the results to the measured spectra.
We discuss our results for the deuteron \pt~distributions as well as for the coalescence parameter $B_{2}$ in comparison to ALICE data in Sec.~\ref{sec:results}.

\section{Wigner function formalism}
\label{sec:wigner}
%%%%%%%%%%%%%%% dedicated to Kachelrieẞ %%%%%%%
As mentioned in Sec.~\ref{sec:introduction}, with the use of Wigner functions, it is possible to describe the production of nuclei via coalescence. In the process of deuteron formation via coalescence, the interactions of the nucleons of the pair with the rest of the particles are assumed to be subdominant due to the low particle density~\cite{Lisa:2005dd}. Hence, the Lorentz-invariant yield of deuterons with momentum $\vec{P}$ can be written as~\footnote{In this derivation, three vectors and four vectors are represented with an arrow ($\vec{a}$) and in italic (\textit{a}), respectively.}
\begin{multline}
    \gamma \, \frac{\mathrm{d} N_{\mathrm{d}}}{\mathrm{d}^3 P}=\frac{S_{\mathrm{d}}}{(2 \pi)^4} \int \mathrm{d}^4 x_1 \int \mathrm{d}^4 x_2 \int \mathrm{d}^4 x_1^{\prime} \int \mathrm{d}^4 x_2^{\prime} \, \times\\
    \times\,\Psi_{\mathrm{d}, P}^*\left(x_1^{\prime}, x_2^{\prime}\right) \Psi_{\mathrm{d}, P}\left(x_1, x_2\right) \rho_{1,2}\left(x_1, x_2 ; x_1^{\prime}, x_2^{\prime}\right) \, , 
\end{multline}
where $\Psi_{\mathrm{d}, P}\left(x_1, x_2\right)$ is the bound-state Bethe-Salpeter amplitude describing the deuteron, $\rho_{1, 2}$ is the density matrix of the two nucleons, and $S_{\mathrm{d}} = 3/8$ is a factor that takes into account spin-isospin statistics.\\
We assume that the two-nucleon density matrix can be factorised into single-nucleon densities 
\begin{equation}
  \rho_{1, 2}\left(x_1, x_2 ; x_1^{\prime}, x_2^{\prime}\right)=\rho_1\left(x_1^{\prime}; x_1 \right) \rho_1\left(x_2^{\prime}; x_2 \right) \, .
\end{equation}
The single nucleon density $\rho_1$ can be written in terms of the single particle Wigner function $f_1^W$~\cite{Blum:2019suo}
\begin{equation}
    \rho_1\left(x, x^{\prime}\right)=\int \frac{\mathrm{d}^4 k}{(2 \pi)^4} \, e^{ik\left(x^{\prime}-x\right)} \, f_1^W\left(k, \frac{x+x^{\prime}}{2}\right) \, .
\end{equation}
Moreover, following the procedure described in \cite{Bellini:2020cbj}, the deuteron Bethe-Salpeter amplitude can be written factoring out the motion of the deuteron

%Defining the coordinates $c_{1,2}=\left(p_{1,2} \, P\right) / P^2$, where the pair total momentum is $P=p_1+p_2 \equiv 2 p$. In general, the dependence of the amplitude $\Psi_{\mathrm{d}, P}$ on the pair total momentum and center of mass coordinate $r_{\mathrm{d}}=c_1 x_1+c_2 x_2$ can be factored out from the dependence on the relative momentum $k=c_2 p_1-c_1 p_2$ and relative position $r=x_1-x_2$~\cite{Bellini:2020cbj}.
\begin{equation}
    \Psi_{\mathrm{d}}\left(x_1, x_2\right) = e^{-i P\cdot r_{\mathrm{d}}} \, \varphi_{\mathrm{d}}(r) \, ,
\end{equation}
where $r_{\mathrm{d}}$ is the space-time position of the deuteron, $P$ its four-momentum, and $\varphi_{\mathrm{d}}(r)$ is the deuteron spatio-temporal wavefunction.\\
Hence, the deuteron spectrum takes the form
\begin{multline}
\gamma \, \frac{\mathrm{d} N_{\mathrm{d}}}{\mathrm{d}^3 P}=  \frac{S_{\mathrm{d}}}{(2 \pi)^4} \int \mathrm{d}^4 r_{\mathrm{d}} \int \frac{\mathrm{d}^4 q}{(2 \pi)^4} \int \mathrm{d}^4 r \, \tilde{\mathcal{D}}(q, r) \, \times \\
 \times \, f_1^W\left(P/2+q,r_{\mathrm{d}}+\frac{r}{2} \right) f_1^W\left(P/2-q,r_{\mathrm{d}}-\frac{r}{2} \right) ,
\end{multline}
where we define the relativistic internal Wigner density as 
\begin{multline}
 \tilde{\mathcal{D}}(q, r)=\int \mathrm{d}^4 \xi \, e^{i q \xi} \, \varphi_{\mathrm{d}}\left(r+\frac{\xi}{2}\right) \varphi_{\mathrm{d}}^*\left(r-\frac{\xi}{2}\right) \, .
\end{multline}
Adapting the Wigner approach used in~\cite{KachelriessLast} to a four-dimensional space leads to 
\begin{multline}
\gamma \, \frac{\mathrm{d} N_{\mathrm{d}}}{\mathrm{d}^3 P}=\frac{S_{\mathrm{d}}}{(2 \pi)^4} \int \mathrm{d}^4 r \int \mathrm{d}^4 r_{\mathrm{d}}  \int \frac{\mathrm{d}^4 q}{(2 \pi)^4} \, \tilde{\mathcal{D}}(q, r) \times \\
\times W_{n p}\left(P / 2+q, P / 2-q, r, r_{\mathrm{d}}\right) \, ,
\label{eq:rel_sepectrum_w}
\end{multline}
where we defined
\begin{multline}
W_{n p}\left(P / 2+q, P / 2-q, r, r_{\mathrm{d}}\right) = \\ f_1^W\left(P/2+q,r_{\mathrm{d}}+\frac{r}{2} \right) f_1^W\left(P/2-q,r_{\mathrm{d}}-\frac{r}{2} \right) \, .
\end{multline}
Using the smoothness and equal-time approximations, as done in \cite{Bellini:2020cbj}, in the pair rest frame (PRF) we obtain $P = (M, \vec{0})$, $q = (0, \vec{q})$ and $ r = (t^*, \vec{r}^*)$, and the Bethe-Salpeter amplitude becomes independent of time in the non-relativistic limit 
\begin{equation}
\Psi\left(q, r\right) \rightarrow \Psi\left(\vec{q}, \vec{r}^*\right).
\end{equation}
Defining $t_1^*$ and $t_2^*$ the emission time of the two nucleons in the PRF, in Eq.~\ref{eq:rel_sepectrum_w} one can write the relative distance of the nucleons as $r = (t_1^*-t_2^*, \vec{r}\,)$ and the center-of-mass coordinate of the nucleons as $r_{\mathrm{d}} = (r_{\mathrm{d}}^0, \vec{r}_{\mathrm{d}})$. Therefore, the integral over the four-momentum $q$ becomes an integral over the momentum $\vec{q}$

\begin{multline}
\gamma \, \frac{\mathrm{d} N_{\mathrm{d}}}{\mathrm{d}^3 P}=\frac{S_{\mathrm{d}}}{(2 \pi)^7} \int \mathrm{d}^4 r \int \mathrm{d}^4 r_{\mathrm{d}} \int \mathrm{d}^3 q \, \mathcal{D}(\vec{q}, \vec{r}) \times\\\times
W_{n p}\left(\vec{P} / 2+\vec{k}, \vec{P} / 2-\vec{q}, r, r_{\mathrm{d}}\right),
\label{eq:long_with_trhee_q}
\end{multline}
where $\mathcal{D}(\vec{q}, \vec{r})$ is the Wigner density in a three-dimensional space. The time of kinetic freeze-out, \mbox{$r_{\mathrm{d}}^0 = t_f$}, represents the moment in which the momentum of the final-state particles is fixed. Separating the space- and time-integrals, Eq.~\ref{eq:long_with_trhee_q} becomes
\begin{multline}
\gamma \, \frac{\mathrm{d} N_{\mathrm{d}}}{\mathrm{d}^3P}=\frac{S_{\mathrm{d}}}{(2 \pi)^7} \int \mathrm{d}^3 r \int \mathrm{d}^3 r_{\mathrm{d}} \, \times \\
\times \, \int dt^*\int dt_f  \int \mathrm{d}^3 q \, \mathcal{D}  (\vec{q}, \vec{r}) \, \times \\
\times \, W_{n p}\left(\vec{P} / 2+\vec{q}, \vec{P} / 2-\vec{q}, \vec{r}, \vec{r}_{\mathrm{d}},t^*,t_f\right) \, .
\label{eq:spatioTemproalYeildD}
\end{multline}
Assuming that $t_f$ is fixed and is the same for all particles, and considering that the particle yield is fixed at a common time (chemical freeze-out), the integral over $t_f$ can be omitted. In addition, due to the time equalisation in the PRF $2\pi\delta(t^*-t_\mathrm{eq})$, choosing arbitrarily $t_\mathrm{eq}=0$ one obtains $t^*=0$. Hence, the integration over $t^*$ in Eq.~\ref{eq:spatioTemproalYeildD} removes the dependence on $t^*$, giving, as a result, a genuine three-dimensional equation~\footnote{From here on, the theoretical framework is constructed in a three-dimensional space. All the vector quantities and their norms are represented with an arrow ($\vec{a}$) and in italic (\textit{a}), respectively, unless specified otherwise.}
\begin{multline}
\gamma \, \frac{\mathrm{d} N_{\mathrm{d}}}{\mathrm{d}^3 P}=\frac{S_{\mathrm{d}}}{(2 \pi)^6} \int \mathrm{d}^3 r \int \mathrm{d}^3 r_{\mathrm{d}} \int \mathrm{d}^3 q \times \\
\times \, \mathcal{D}(\vec{q}, \vec{r}) \,  
W_{n p}\left(\vec{P} / 2+\vec{q}, \vec{P} / 2-\vec{q}, \vec{r}\,,\vec{r_{\mathrm{d}}}\right),
\label{eq:spatialYeildD}
\end{multline}
The three-dimensional Wigner function of the deuteron $\mathcal{D}(\vec{q}, \vec{r})$ is defined as
\begin{equation}
  \mathcal{D}(\vec{q}, \vec{r}) = \int \mathrm{d}^{3}\xi \, e^{-i \, \vec{q} \cdot\vec{\xi}} \, \varphi_{\mathrm{d}}(\vec{r} + \vec{\xi}/2) \,  \varphi^{*}_{\mathrm{d}}(\vec{r} - \vec{\xi}/2).
  \label{eq:deuteron_wigner}
\end{equation}
Notably, the choice of the deuteron wavefunction $\varphi_{\mathrm{d}}$ affects only the term $\mathcal{D}(\vec{q}, \vec{r})$ in Eq.~\ref{eq:spatialYeildD}, while the other terms remain the same. \\
The starting point in this theoretical derivation are the single free nucleon momentum distributions. In principle, there is no overlap between the deuteron Wigner function ($\mathcal{D}(\vec{q}, \vec{r})$) and the free nucleon one ($W_{n p}$). This problem arises from the conservation of energy and it can be resolved by introducing a third particle (usually a pion), see e.g. Ref.~\cite{Scheibl:1998tk}. However, as done in previous works~\cite{Kachelriess:2020uoh,KachelriessLast,Scheibl:1998tk}, we make a semi-classical approximation where we assume the binding energy to be only a negligible correction, as it is much smaller than the mass scale of the nucleons (2.2 MeV $\sim E_B\ll m_p =$ 0.938 GeV).

In~\cite{KachelriessLast} a factorisation of space and momentum dependence of the proton--neutron Wigner function is assumed

\begin{equation}
  W_{\mathrm{np}} = H_{\mathrm{np}}(\vec{r}_{\mathrm{n}}, \vec{r}_{\mathrm{p}}) \, G_{\mathrm{np}}(\vec{P}_{\mathrm{d}}/2 + \vec{q}, \vec{P}_{\mathrm{d}}/2 - \vec{q}),
  \label{eq:pn_wigner}
\end{equation}
where $G_{\mathrm{np}}$ is the two-particle momentum distribution, taken from MC generators, containing the nucleon single-particle momentum distributions and their initial-state correlation.
Furthermore, for the space term $H_{\mathrm{np}}$, the spatial correlation is neglected,

\begin{equation}
  H_{\mathrm{np}}(\vec{r}_{\mathrm{p}}, \vec{r}_{\mathrm{n}}) = h(\vec{r}_{\mathrm{p}}) \, h(\vec{r}_{\mathrm{n}}),
  \label{eq:factorisation}
\end{equation}
and the spatial single-particle distributions $h(\vec{r}_{p,n})$ are assumed to be Gaussian, hence

\begin{equation}
  H_{\mathrm{np}}(\vec{r}, \vec{r}_{\mathrm{d}}; r_{0}) = \frac{1}{\left(2 \pi \, r_{0}^2\right)^{3}} \, \exp{\left(-\frac{r^{2}+r_\mathrm{d}^2}{4 \, r_{0}^2}\right)}.
  \label{eq:space_corr}
\end{equation}
Here, $\vec{r}_\mathrm{d}\equiv\vec{r}_\mathrm{p}+\vec{r}_\mathrm{n}$, $r_{0}$ is the size of the two-particle emitting source, and $\vec{r}\equiv\vec{r}_\mathrm{p}-\vec{r}_\mathrm{n}$ as before. 
In our work, possible space-momentum correlations and two-particle correlations in momentum and space coordinates at the hadron production point are considered in the source model discussed in Sec.~\ref{sec:source}.
Finally, the coalescence probability~\footnote{Formally, this is not a probability, as $\mathcal{P}(r_{0}, q) \in [0, 8]$. This is due to the quantum-mechanical nature of the process and to the definition of the Wigner function. Indeed, $\mathcal{D}(\vec{0}, \vec{0}) =$ 8, regardless of the wavefunction.} $\mathcal{P}(r_{0}, q)$ can be obtained by folding the spatial distribution of nucleons with the deuteron Wigner function

\begin{equation}
    \mathcal{P}(r_{0}, q)= \int d^3r_\mathrm{d}\int d^3r \, H_{\mathrm{pn}}(\vec{r}, \vec{r}_{\mathrm{d}}; r_{0}) \,\,\mathcal{D}(\vec{q}, \vec{r}) \,   . 
    \label{eq:prob}
\end{equation}
Equations~\ref{eq:space_corr} and \ref{eq:prob} are based on the assumption of a Gaussian source, for which the mean value ($r_{\mu}$) of the two-particle distance distribution is related to the source size $r_{0}$ through $r_{\mu} = (4/\sqrt{\pi}) \, r_{0}$. However, in particle generators the source has a shape that is generally not Gaussian~\cite{sourceSizeHMpp}. For this reason, we evaluate $r_{\mu}$ from a fit to the distribution and then we obtain the source size $r_{0}$ through their relation. %$r_{0} = (\sqrt{\pi}/4) \, r_{\mu}$.
After this derivation, the deuteron momentum distribution in Eq.~\ref{eq:spatialYeildD} assumes the final form

\begin{equation}
  \frac{\mathrm{d}^{3}N_{\mathrm{d}}}{\mathrm{d}P_{\mathrm{d}}^{3}} = S_\mathrm{d} \int d^3q \, \mathcal{P}(r_{0}, q) \frac{G_{\mathrm{np}}(\vec{P}_{\mathrm{d}}/2 + \vec{q}, \vec{P}_{\mathrm{d}}/2 - \vec{q})}{(2\pi)^6}.
  \label{eq:deuteron_spectrum_wigner_final}
\end{equation}

\section{Source}
\label{sec:source}
To account for realistic particle emission and correlation, in our implementation of the coalescence afterburner we rely on the ALICE measurement of the nucleon-emitting source~\cite{sourceSizeHMpp}, which has been performed differentially in transverse mass only in pp collisions at \mbox{$\sqrt{s} = 13$~TeV}. As mentioned before, a measurement of the deuteron production spectra~\cite{nuclei_pp_13TeV_HM}, which we use to validate our model, is available in the same system as discussed in Sec.~\ref{sec:results}.

To simulate pp events we employ two different MC generators, EPOS 3.117 and PYTHIA 8.3 with the Monash 2013 tune configuration. 
\mbox{EPOS 3}~\cite{Werner:2013tya} is a hybrid Monte-Carlo event generator
in which the reaction volume is divided into ``core'' and ``corona'', depending on the local density and transverse momentum of the string segments. The core represents a thermalised bulk of matter evolving according to 3+1D viscous hydrodynamics and hadronises according to a Cooper-Frye mechanism~\cite{Cooper:1974mv,PhysRevD.11.192}. The particles in the corona originate from string fragmentation. As a function of final-state particle multiplicity, the relative fraction of core and corona evolves dynamically. 
On the other hand, PYTHIA 8.3~\cite{Sjostrand:2014zea} is a parton-based microscopic event generator, where the primary process of a pp collision is represented with hard parton scatterings via 2 $\rightarrow$ 2 matrix elements defined at leading order. It is complemented by parton showers that include initial- and final-state radiation via the leading-logarithm approximation. 
The hadronisation from partons is performed using the Lund string fragmentation model~\cite{Andersson:1983ia}. 
The Monash 2013 tune~\cite{Skands:2014pea} is chosen because it improves the description of minimum-bias and underlying-event observables in pp collisions at LHC energies and it includes a multi-parton interaction-based color-reconnection scheme. The underlying event in the model consists of particles  originating from multi-parton interactions as well as from beam remnants. In the color reconnection picture, the strings between partons can be rearranged in a way that the total string length is reduced.

A proper modelling of the nucleon-emitting source, needs to take into account a) the overall final-state particle multiplicity, b) the possible contribution of feed-down from strongly-decaying resonances, and c) the phase-space correlations among the particles of interest.

Starting from point a), the MC simulations are required to reproduce the average charged-particle multiplicity density at midrapidity as measured.
The considered ALICE data were collected with a high-multiplicity trigger, that corresponds to \mbox{0--0.01\%} of the total inelastic pp cross section.  %$\sigma^{pp}_{\mathrm{INEL}}$. 
Following the nomenclature used in~\cite{nuclei_pp_13TeV_HM}, this multiplicity class is referred to as \mbox{\textit{HM I}}. 
In these events, the average multiplicity density is $\langle$d$N_{\mathrm{ch}}/$d$\eta\rangle_{\vert\eta\vert < 0.5} = 35.8 \pm 0.5$~\cite{nuclei_pp_13TeV_HM}. 
Therefore, the simulated events of interest are selected (triggered) based on the number of charged particles produced at backward and forward rapidity, mimicking the selection method of ALICE~\cite{nuclei_pp_13TeV_HM}.
The trigger acceptance factors are $10^{-2}$ for EPOS 3 and $2\times 10^{-5}$ for PYTHIA 8.3. 

Regarding b), as described in detail in~\cite{sourceSizeHMpp}, short-lived resonances that decay into protons and neutrons, the particles of interest, play a crucial role in the determination of the source size. Indeed, the emitting source can be modeled as the convolution of a Gaussian core, which is the same for all baryons~\cite{sourceSizeHMpp}, and an exponential tail related to the decay of resonances. This means that a wrong resonance cocktail will influence the overall source size.
For this reason, the relative fractions of the resonance cocktail of the MC simulations are tuned using a statistical hadronisation model, namely ThermalFIST~\cite{TheFIST}. As input to the model, the measured temperature of chemical freeze-out and the correlation volume~\cite{Vovchenko:2019kes} corresponding to the \mbox{\textit{HM I}} class are provided.  

The source size $r_{0}$ obtained directly from the event generators after the tuning of the resonance cocktail does not match the ALICE measurement~\cite{sourceSizeHMpp} and hence it needs to be calibrated event-by-event. 
We reproduce the measured source size by acting at the particle-propagation level in the event generators, allowing for the conservation of the phase-space correlations provided by the event generator. 
In this regard, our approach differs from that used in~\cite{KachelriessLast}, which employs an analytical form for $r_{0}$, fitted to the radii measured by ALICE. 

\begin{figure}
    \centering
    \includegraphics[width=0.99\linewidth]{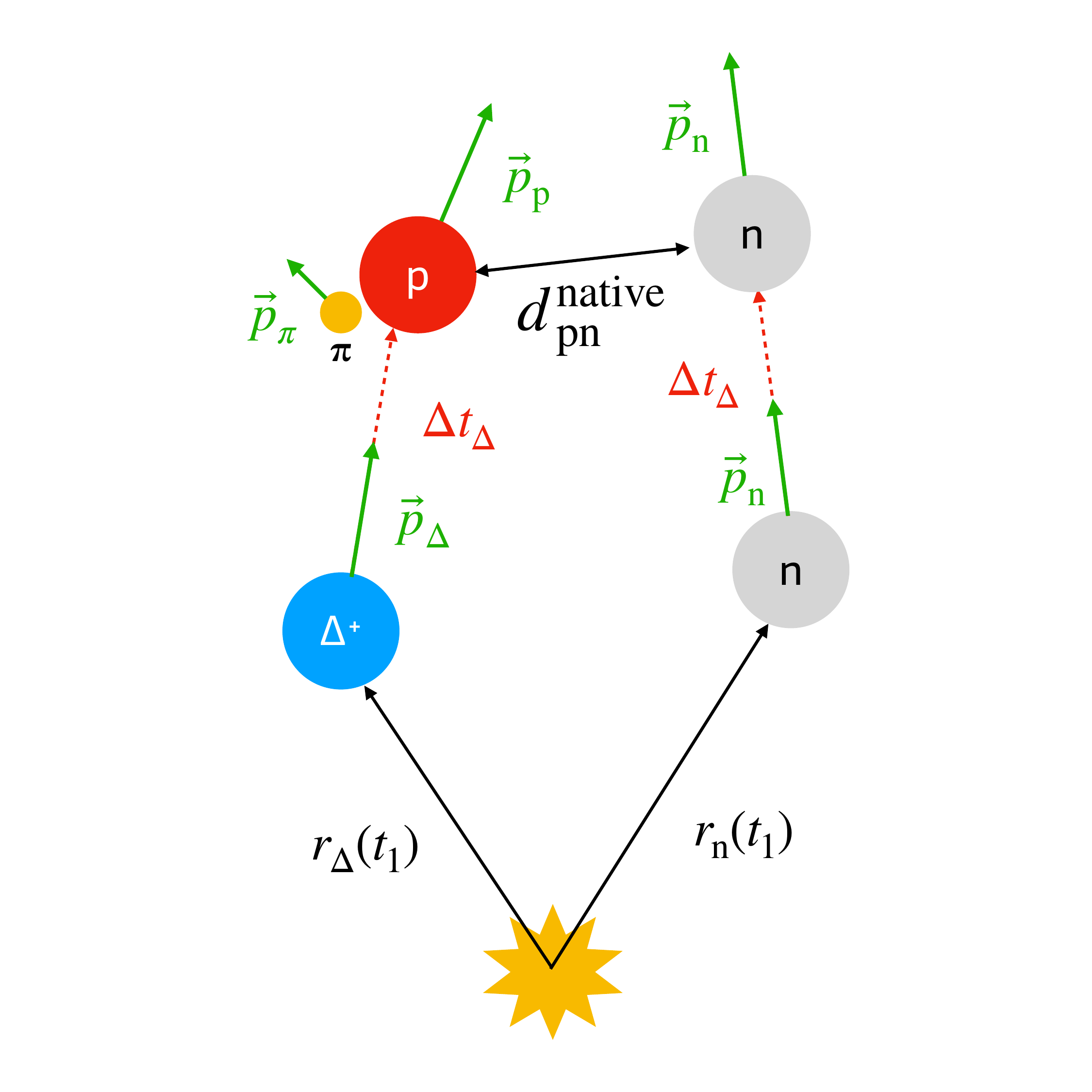}
    \caption{Scheme of the determination of the proton-neutron distance in MC simulations 
    std::vector<int> Mults{1, 2, 3, 4, 5, 6, 7, 9, 12, 15, 18, 22, 25, 28, 31, 34, 40, 50, 70, 90};$d_\mathrm{pn}^\mathrm{native}$ in the case of a proton coming from the decay of a $\Delta^+$ resonance. See text for more details. }
    \label{fig:ScaledSourceComic_Model}
\end{figure}

\begin{figure}
    \centering
    \includegraphics[width=0.99\linewidth]{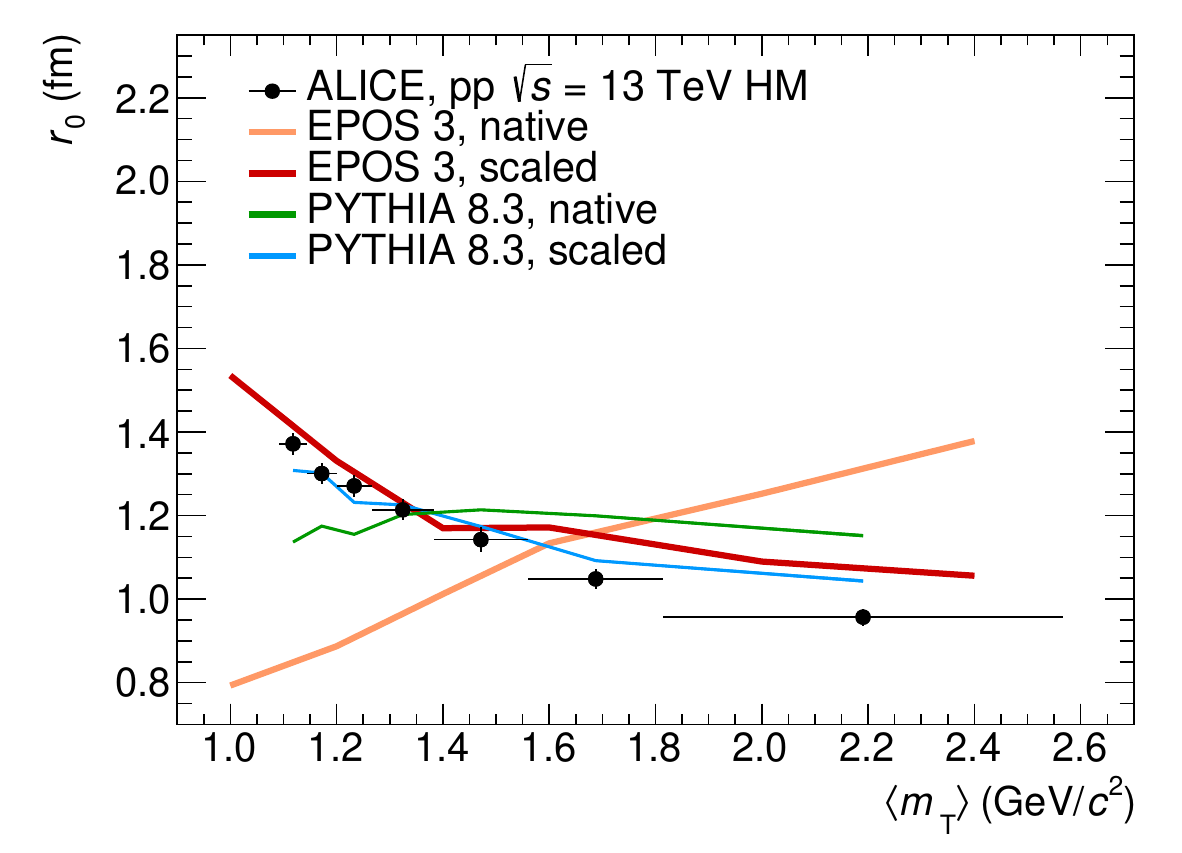}
    \caption{Comparison between the source size $r_{0}$ measured by ALICE~\cite{sourceSizeHMpp}, the native ones for EPOS 3 (orange) and PYTHIA 8.3 (green) and those obtained after the source modelling (in red and blue for EPOS 3 and PYTHIA 8.3, respectively), as a function of the average transverse mass $\langle m_{\mathrm T}\rangle$ of the proton-proton (antiproton-antiproton) pair. For the ALICE data, statistical and systematic uncertainties are summed in quadrature and shown as vertical bars, while for EPOS 3 and PYTHIA 8.3 uncertainties are negligible.}
    \label{fig:rvsmTBefore}
\end{figure}

The starting point of our source model are the space-time coordinates of the nucleons produced by the event generator.  Figure~\ref{fig:ScaledSourceComic_Model} illustrates an example of the implementation for a primordial neutron paired with a proton stemming from a $\Delta^+$ resonance. To take into account that the particles in this pair are not necessarily created at the same time, the particle created earlier is propagated along its momentum for the time difference between the two. In Fig.~\ref{fig:ScaledSourceComic_Model}, the neutron and the $\Delta^+$ resonance are depicted at time $t_1$ at distances $r_{\rm n}(t_1)$ and $r_{\Delta}(t_1)$ from the production point, respectively, after the aforementioned time equalisation. The resonance decays after a time interval $\Delta t_{\Delta}$, during which the neutron moves. To estimate the distance between the neutron and the final-state proton, the neutron is propagated along its momentum for the time $\Delta t_{\Delta}$. The distance between the proton and the neutron $d_\mathrm{pn}^\mathrm{native}$ is evaluated at the time $t_1 +\Delta t_{\Delta}$. 
In the case in which both nucleons come from resonances, the one with the smallest time component is propagated along its momentum until the time equalisation is achieved. Finally, if a resonance decays into the particle of interest in a multi-step process (e.g. as it is the case for $\Delta(1900)^{++} \rightarrow N^*(1440)^+ \rightarrow p$), $\Delta t_{\Delta}$ is defined as the time difference between the last decay and the production of the first resonance.
After the resonances decay and the final state particles have equal time ($t_1 +\Delta t_{\Delta}$), the distance $d_\mathrm{pn}^\mathrm{native}$, in the proton-neutron pair rest frame, and the average transverse mass $\langle m_{\mathrm{T}}\rangle$ of the resulting pair~\footnote{The average transverse mass of a pair of two particles with mass $m_{1}$ and $m_{2}$ is here defined as 
\begin{equation*}
\langle m_{\mathrm{T}}\rangle = \sqrt{\bigg(\frac{p_\mathrm{T,1} + p_\mathrm{T,2}}{2}\bigg)^{2} + \bigg(\frac{m_{1} + m_{2}}{2}\bigg)^{2}}.
\end{equation*}} 
are stored. 
The native source size $r_{0}^{\mathrm{native}}$ is obtained from the mean value $r_{\mu}^{\mathrm{native}}$ of the $d_\mathrm{pn}^\mathrm{native}$ distribution as \mbox{$r_{0}^{\mathrm{native}} = (\sqrt{\pi}/4) \, r_{\mu}^{\mathrm{native}}$} (see Sec.~\ref{sec:wigner}).
In Fig.~\ref{fig:rvsmTBefore}, $r_{0}^{\mathrm{native}}$ is shown as a function of $\langle m_{\mathrm{T}}\rangle$ of the particle pair for both EPOS 3 (orange line) and PYTHIA 8.3 (green line) simulations. 
From the comparison with the ALICE measurement $r_{0}^\mathrm{ALICE}$ ~\cite{sourceSizeHMpp} (black points), it is clear that the native source size predicted by MC generators does not reproduce the observed $m_{\mathrm{T}}$-dependence. Hence, an additional $m_{\mathrm{T}}$-scaling $\mathcal{S}(m_{\mathrm{T}}) = r_{0}^\mathrm{ALICE} / r_{0}^\mathrm{native}$ is introduced and the corrected proton-neutron distance is obtained as \mbox{$d_{\mathrm{pn}}^\mathrm{scaled} = \mathcal{S}(m_{\mathrm{T}}) \, d_{\mathrm{pn}}^\mathrm{native}$}.
\\In summary, with this model, we succeed in preserving the space-momentum correlations among the nucleons, which were explicitly broken in the factorisation shown in Eq.~\ref{eq:factorisation}, by reproducing the measured source size. This is shown in Fig.~\ref{fig:rvsmTBefore} for both EPOS 3 (red line) and PYTHIA 8.3 (blue line).
\\Finally, the angular correlations of the nucleon pairs in the event generators do not match the $\Delta\phi$-$\Delta\eta$ correlations measured by ALICE~\cite{DphiDeta7TeVMB}. However, reweighting them would destroy the nucleon space-momentum correlations provided by the event generators. More details on the angular correlations are given in Appendix~\ref{app:AngularCorrelation}.

\section{Results}
\label{sec:results}
\begin{figure}
    \centering
    \includegraphics[width=0.45\textwidth]{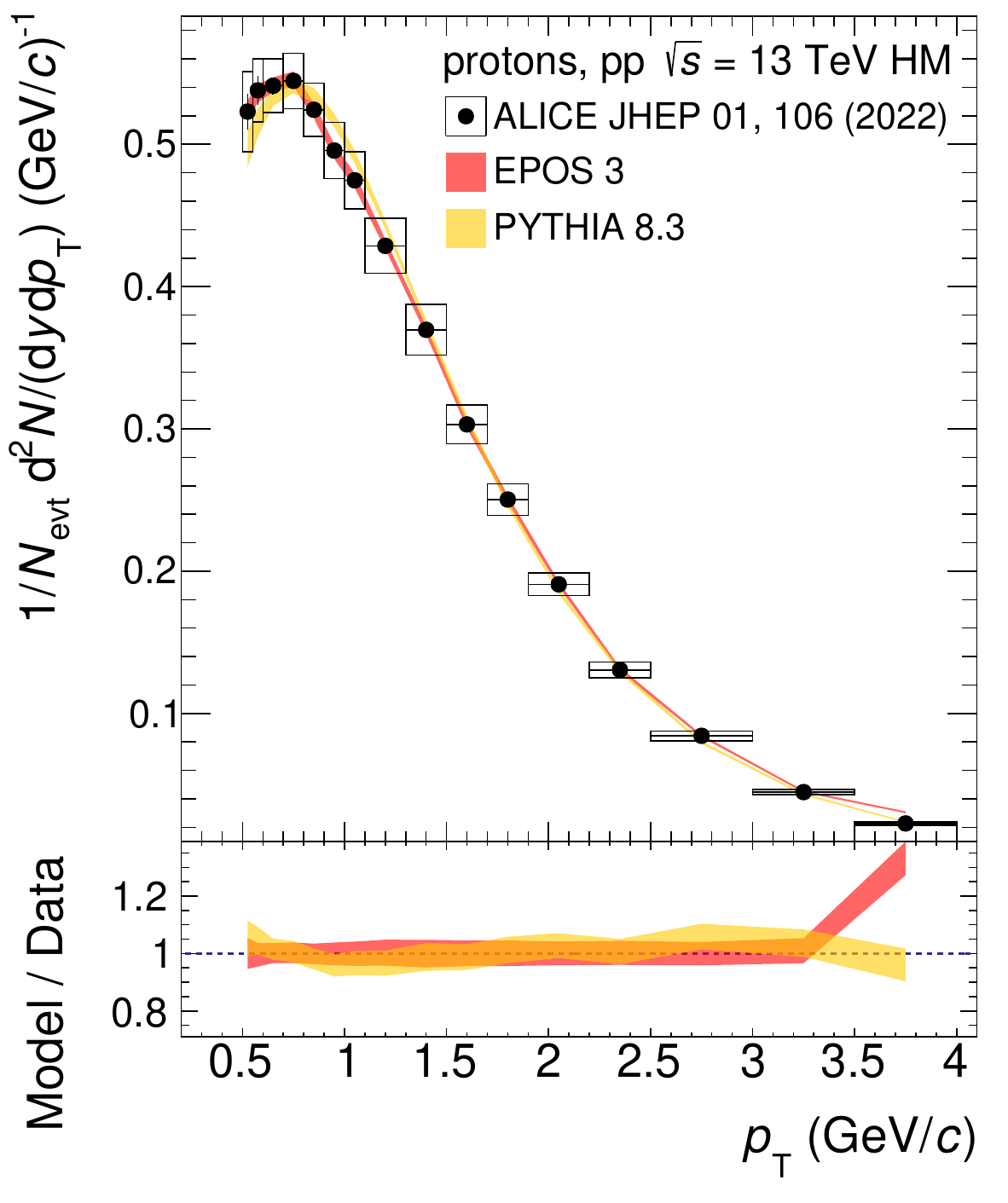}
    \caption{Proton spectra generated by EPOS 3 and PYTHIA 8.3, compared with the ALICE measurement~\cite{nuclei_pp_13TeV_HM}. In the bottom panel, the data-to-model ratios are shown.}
    \label{fig:proton_spectra}
\end{figure}

The source modelling and the Wigner function formalism described in the previous sections are employed to obtain the deuteron spectra starting from nucleons simulated with EPOS 3 and PYTHIA 8.3.
The proton and neutron~\footnote{Neutrons and protons are assumed to have the same production spectra, since both belong to the same isospin doublet.} spectra are reweighted such to match the proton measurement by ALICE~\cite{nuclei_pp_13TeV_HM} (see Fig.~\ref{fig:proton_spectra}) in order to start from realistic transverse momentum distributions. 

\begin{figure}
    \centering
    \includegraphics[width=0.45\textwidth]{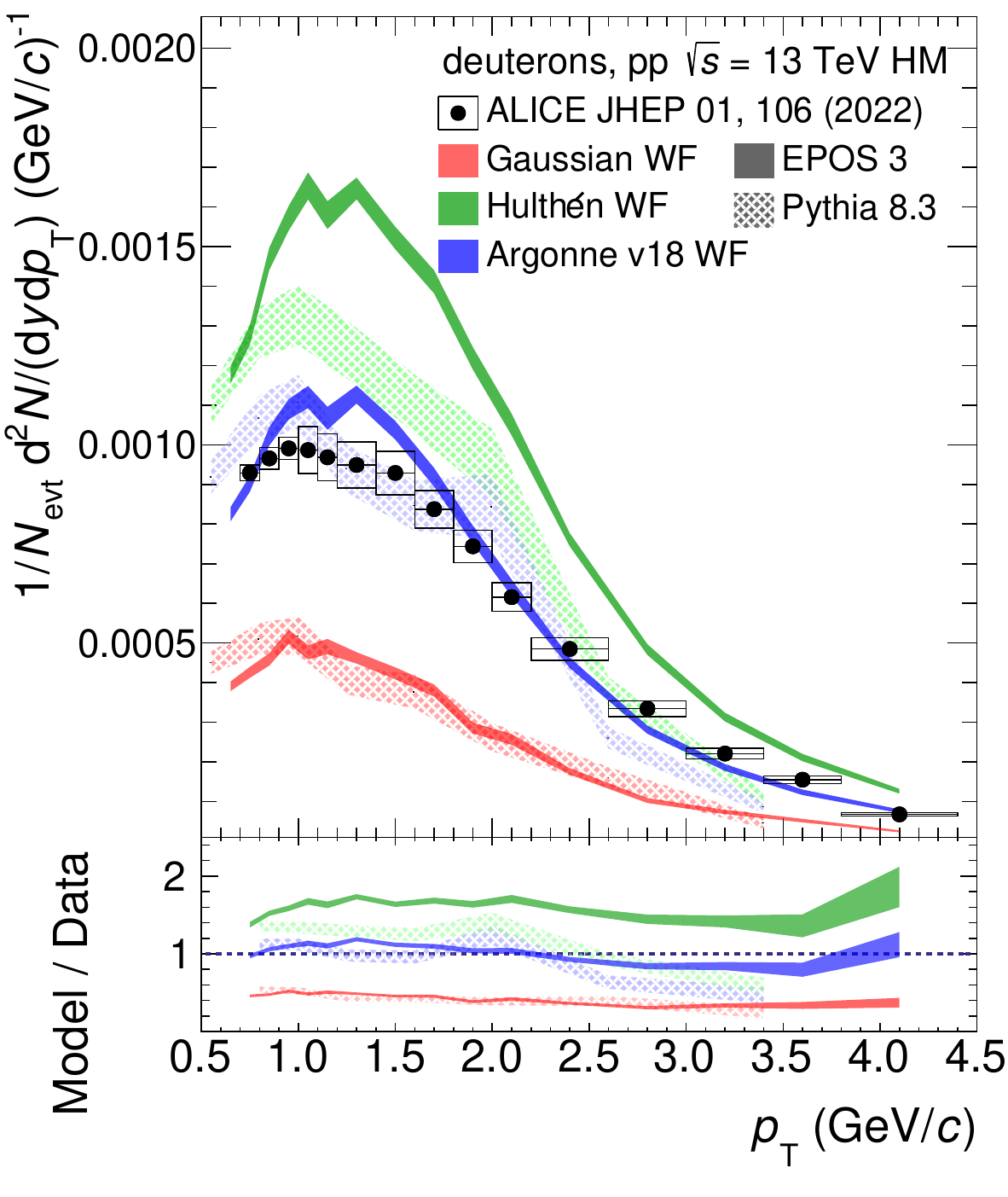}
    \caption{Deuteron spectra obtained from EPOS 3 and PYTHIA 8.3, applying the coalescence model with different hypotheses for the deuteron wavefunction, compared with the ALICE measurement~\cite{nuclei_pp_13TeV_HM}. The width of the bands represents the statistical uncertainty of the models. The systematic uncertainty of the spectra (6\%) is not shown. In the bottom panel, the data-to-model ratios are shown.}
    \label{fig:deuteron_spectra}
\end{figure}
Starting from the generated protons and neutrons, the event-by-event coalescence is implemented as a statistical rejection method. 
For each pair, we apply the source modelling described in Sec.~\ref{sec:source} and calculate the coalescence probability for the single proton-neutron pair\\ \mbox{$\mathcal{P}\left(\frac{\sqrt{\pi}}{4} \, d_{\mathrm{pn}}^{scaled}, q\right)$} using Eq.~\ref{eq:prob}. 
The exact form of \mbox{$\mathcal{P}$} depends on the deuteron wavefunction. For this study, we have considered three different wavefunctions: a Gaussian, Hulth\'en~\cite{Heinz:1999rw}, and Argonne $v_{18}$~\cite{PhysRevC.51.38}. Whereas the Gaussian is the simplest functional form of the wavefunction, the Hulth\'en and Argonne $v_{18}$ are based on physical properties of the deuteron. While the Argonne $v_{18}$ is expected to yield the best results, as it reproduces modern scattering data with a $\chi^2\sim 1$, the Hulth\'en and the Gaussian are merely considered for comparison with previous studies.
The shape of the deuteron wavefunction in the asymptotic region (\mbox{r $\gtrsim 1.5$ fm}) is determined by the solution of the Schr\"{o}dinger equation for an interaction potential that reproduces the deuteron binding energy correctly ($E_{\mathrm{B}}\sim$ 2.2 MeV). In this range, Hulth\'en and Argonne $v_{18}$ are very similar, as shown in Fig.~\ref{fig:wavefunctions}. In the range \mbox{r $\lesssim 1.5$ fm}, the Hulth\'en and Argonne $v_{18}$ wavefunctions drastically differ, as a consequence of the different p-n potentials. Indeed, the Hulth\'en potential corresponds to an attractive interaction, while the Argonne $v_{18}$ contains a repulsive core in the interaction.
A difference in the predicted yields computed with the two wavefunctions suggests that the nuclear production mechanism is sensitive to the short-range strong interaction between nucleons. 
%Further studies using different interaction potentials could lead to a deeper understanding of the deuteron formation mechanism and are deferred to a future publication. 
In this work, the Wigner functions of the Hulth\'en and Argonne $v_{18}$ wavefunctions are computed for the first time, with details given in Appendix~\ref{app:theory}. 
Figure~\ref{fig:deuteron_spectra} shows the deuteron spectra obtained with the different wavefunctions and different event generators. The results using the Argonne $v_{18}$ wavefunction are in excellent agreement with the data measured by ALICE, regardless of the event generator used. This proves that with our model, given the correct source size, nucleon spectra and average charged-particle multiplicity density, and a realistic wavefunction, it is possible to predict deuteron yields accurately.
\\The systematic uncertainties of the model on the final deuteron spectra are estimated to be around 6\%, independent of $p_\mathrm{T}$. The first source of systematic uncertainties is related to the source size. For this, we varied the source size used in the model by $\pm$7\%, based on the uncertainties reported in~\cite{sourceSizeHMpp}. The resulting systematic uncertainty is obtained from the relative deviation in the final spectra between the default source size and the varied one. The second source of systematic uncertainties is related to the fraction of primordial nucleons. To account for this, the primordial nucleon fraction is varied by $\pm$10\% and the relative deviation is the final spectra is considered. The two sources of uncertainties are summed in quadrature.

\begin{figure}[t]
    \centering
    \includegraphics[width=0.45\textwidth]{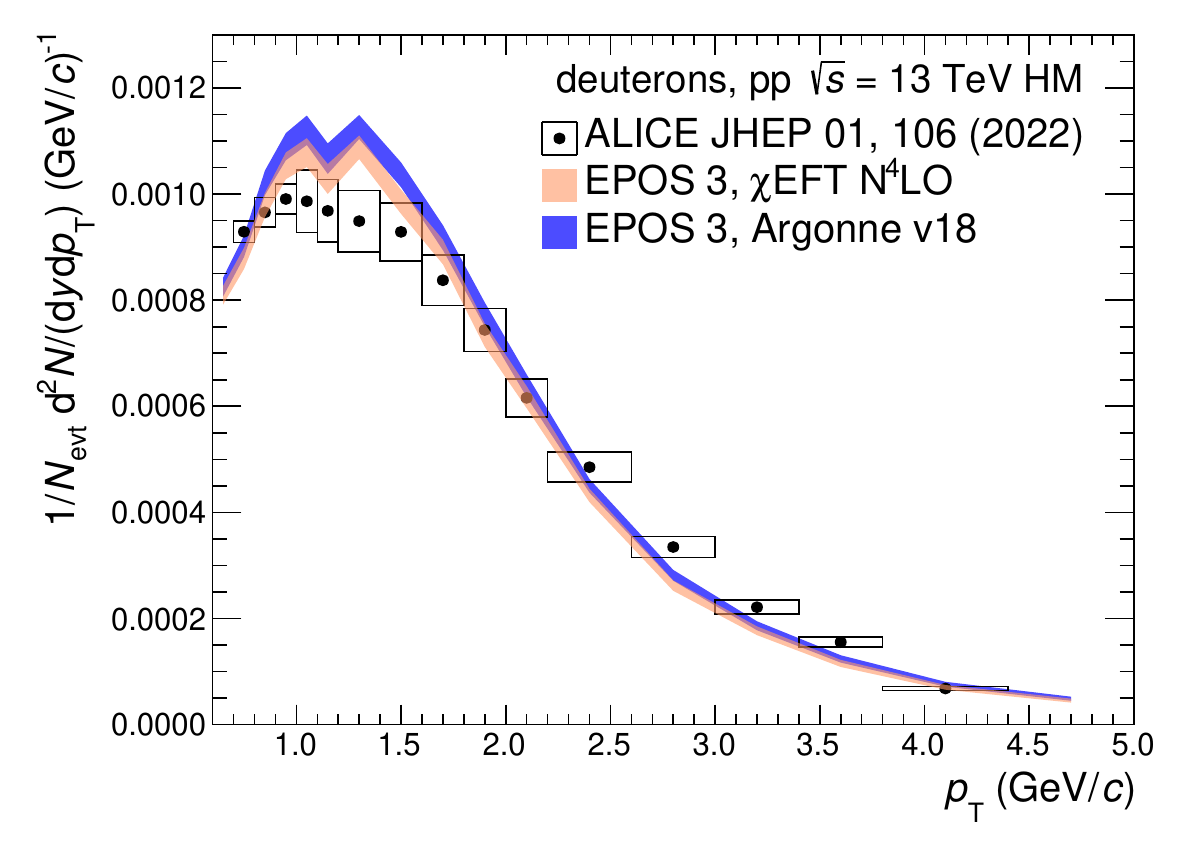}
    \caption{Deuteron spectra obtained with EPOS 3, comparing two wavefunction hypotheses, i.e. Argonne $v_{18}$ and $\chi$EFT. Predictions are compared with the corresponding ALICE measurement~\cite{nuclei_pp_13TeV_HM}.}
    \label{fig:deuteron_ChiEFT}
\end{figure}

In order to further test the impact of the core part of the strong-interaction potential on the deuteron yield, the deuteron wavefunction obtained from \textit{ab initio} Chiral Effective Field Theory ($\chi$EFT)~\cite{chiEFT} (N$^4$LO) calculations are employed to compute the deuteron yield and the results are compared with those obtained using the Argonne $v_{18}$ wavefunction (see Fig.~\ref{fig:deuteron_ChiEFT}). As for the Hulth\'en and Argonne $v_{18}$ wavefunctions, also the $\chi$EFT one is computed for the first time in this work, with details reported in Appendix~\ref{app:theory}.
On the one hand, the Argonne $v_{18}$ potential constructs the core part of the interaction using a combination of central, tensor, and spin-orbit interactions. On the other hand, the $\chi$EFT NN potential is derived systematically from the underlying theory of QCD and involves a perturbative expansion in powers of a small parameter related to the pion mass. The two-body interactions in $\chi$EFT are calculated using Feynman diagrams that involve pions and nucleons. The leading order (LO) two-body interactions involve only one-pion exchange, while higher-order (NLO, NNLO, etc.) interactions involve multiple pion exchanges and nucleon self-interactions.\\
While Argonne $v_{18}$ and $\chi$EFT are two different approaches, the deuteron wavefunctions obtained from these approaches do not differ significantly, except for very short ranges. Indeed, the deuteron wavefunction from $\chi$EFT shows less repulsion than the one obtained from the Argonne $v_{18}$ potential. Nevertheless, the predicted deuteron yields using Argonne $v_{18}$ and $\chi$EFT are compatible, indicating that the production of the deuteron is not affected by extremely short-range interactions, as shown in Fig.~\ref{fig:deuteron_ChiEFT}.

\begin{figure}[t]
    \centering
    \includegraphics[width=0.45\textwidth]{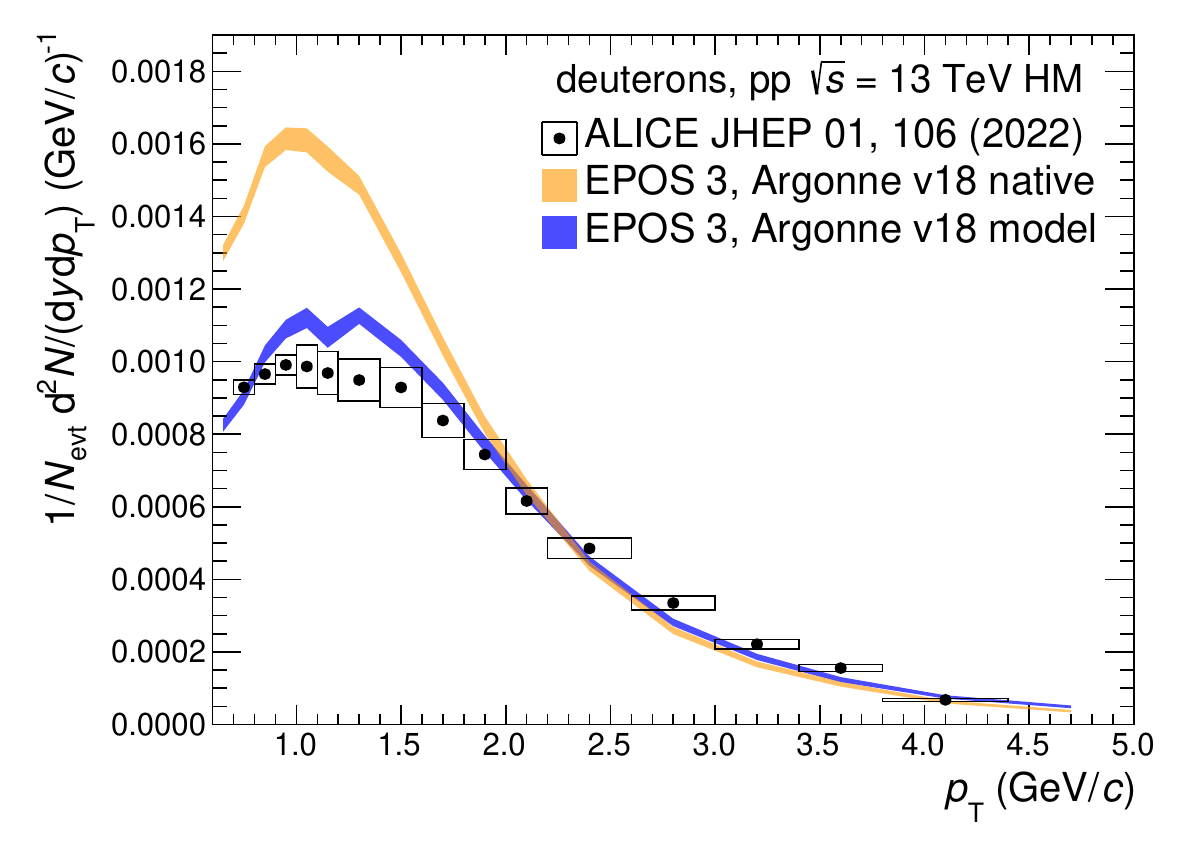}
    \caption{Deuteron spectra obtained with EPOS 3, with the source modelling (model) and without (native), for the same wavefunction hypothesis, i.e. Argonne $v_{18}$. Predictions are compared with the corresponding ALICE measurement~\cite{nuclei_pp_13TeV_HM}.}
    \label{fig:deuteron_wandwo_model}
\end{figure}

Figure~\ref{fig:deuteron_wandwo_model} shows the deuteron spectrum using the Argonne $v_{18}$ wavefunction, with and without the re-modelling of the source. Only the modeled source is able to describe the data. The difference between the two predictions is up to a factor of two, depending on $p_{\rm T}$.

 The impact of the correlation between the relative momentum \textit{q} and the distance between particles \textit{r} taken from the event generator, shown in Fig.~\ref{fig:QRCorrelation}, on the final results is evaluated to be around $10\%$. For this, we compare the deuteron spectrum shown in Fig.~\ref{fig:deuteron_spectra} with one obtained by randomly sampling distances from the source size measurement shown in Fig.~\ref{fig:rvsmTBefore}. Using the $m_\mathrm{T}$-dependent parameterisation of the source size, the final spectra change by around $30\%$ with respect to the ones obtained with the native source provided by the generator, as shown in Fig.~\ref{fig:deuteron_wandwo_model}. Lastly, the difference between the results obtained using a realistic wavefunction and a Gaussian one is around $50\%$.
 
\begin{figure}
    \centering
    \includegraphics[width=0.45\textwidth]{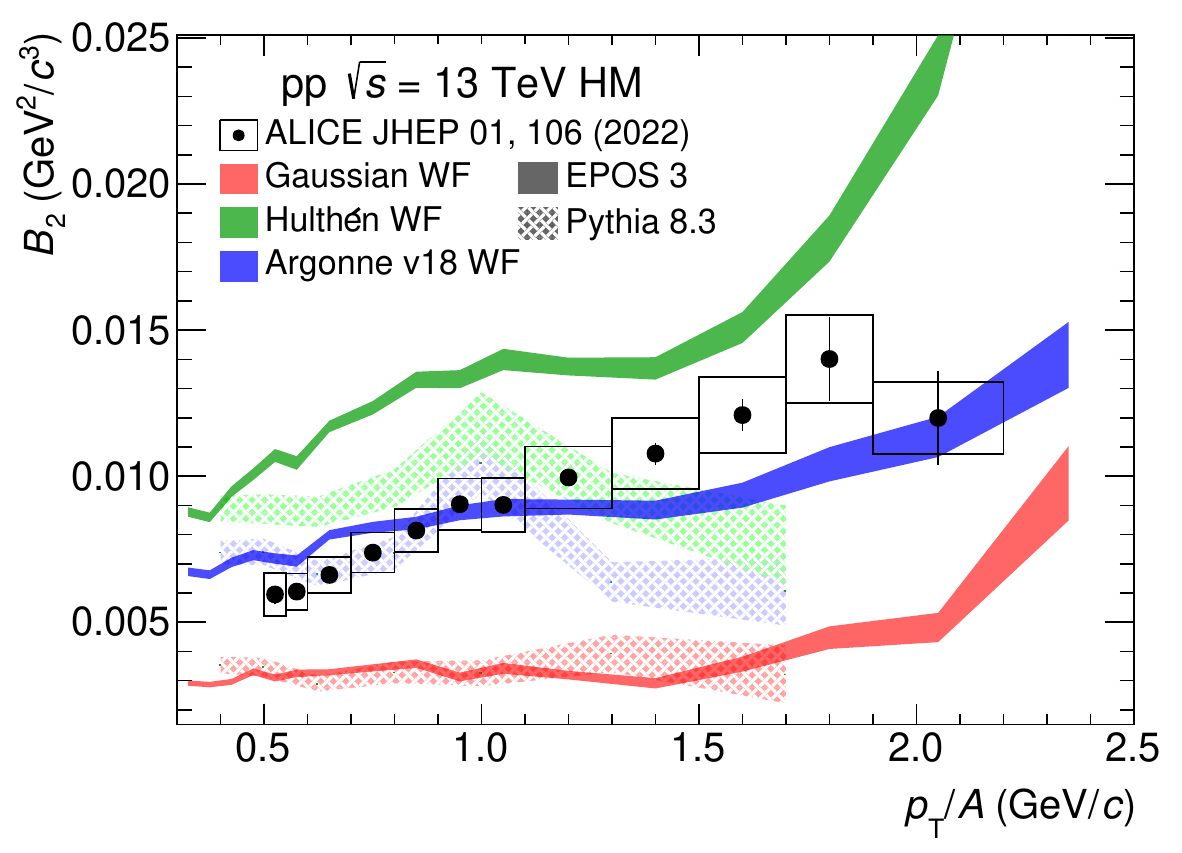}
    \caption{Coalescence parameter $B_2$ obtained with EPOS 3 and Pythia 8.3, compared with the ALICE measurement~\cite{nuclei_pp_13TeV_HM}. }
    \label{fig:B2}
\end{figure}

Using the spectra of protons and deuterons shown in Figs.~\ref{fig:proton_spectra} and~\ref{fig:deuteron_spectra}, it is possible to compute the coalescence parameter $B_2$, as 

\begin{equation}
B_{2} = { \biggl( \dfrac{1}{2 \pi p^{\mathrm d}_{\mathrm T}} \dfrac{ \mathrm{d}^2N_{\mathrm d}}{\mathrm{d}y\mathrm{d} p_{\mathrm T}^{\mathrm d}}  \biggr)}  \bigg/{  \biggl( \dfrac{1}{2 \pi p^{\rm p}_{\mathrm T}} \dfrac{\mathrm{d}^2N_{\mathrm p} }{\mathrm{d}y\mathrm{d}p_{\mathrm T}^{\mathrm p}} \biggr)^2}  .
\label{eq:BA}
\end{equation}
The labels d and p indicate the deuteron and the proton, respectively, and the transverse momentum of protons is half of that of deuterons (\mbox{$p_{\mathrm T}^{\rm p}$ = $p_{\mathrm T}^{\mathrm d}$/$2$}). 
The comparison among the $B_2$ measured by ALICE~\cite{nuclei_pp_13TeV_HM} and those obtained using EPOS 3 and PYTHIA 8.3, is shown in Fig.~\ref{fig:B2}. 
A similar comparison in terms of $B_2$ was done in~\cite{nuclei_pp_13TeV_HM}, where the coalescence predictions were obtained using the formalism described in~\cite{Blum:2019suo}. In that case, the Gaussian wavefunction provided the best description of data, while Hulth\'en overestimated the measurement by a factor of two. In comparison, in this work, the predictions using a Gaussian wavefunction underestimate the coalescence parameter by about 50 to 70\%, while the predictions using the Hulth\'en wavefunction overestimate the measured $B_2$ by about 20 to 50\%. 
Among different assumptions, the main difference between the coalescence predictions shown in~\cite{nuclei_pp_13TeV_HM} and in the present work is that in the formalism described in~\cite{Blum:2019suo} the difference in momentum between the nucleons ($\vec{q} = (\vec{p}_{\mathrm{p}} - \vec{p}_{\mathrm{n}})/2$) is neglected in the Wigner function of the p-n state.  
The authors of~\cite{Blum:2019suo} state that such approximation, motivated by the ease of calculations, is valid to an accuracy of around 10\% in Pb--Pb collisions, while the accuracy in pp collisions is not estimated and could potentially be much larger. %However, neglecting $\vec{q}$ explicitly destroys the space-momentum correlations in Eq.~\ref{eq:spatialYeildD}, and leads to a difference of a factor of about 50\% in the resulting coalescence parameter obtained using the Gaussian wavefunction. 

%in our approach we take into account proton-neutron space-momentum correlations by using event generators, while 

\section{Conclusions}
\label{sec:conclusion}

In this paper, we show the implementation of a coalescence afterburner based on a state-of-the-art Wigner function formalism and use it to reproduce the (anti)deuteron spectra measured by ALICE in pp collisions at $\sqrt{s}=$ 13 TeV, collected with a high-multiplicity trigger. The novelty of our work consists in the preservation of the space-momentum correlation of nucleons, obtained by correcting the $m_\mathrm{T}$ scaling of the source in the event generators with a parameterisation anchored to experimental measurements. At the same time, the Wigner function formalism with realistic deuteron wave functions is employed. The constraint to the measured source size allows for an accurate prediction of the (anti)deuteron spectra. In this work, three different hypotheses for the internal wavefunction of the deuteron are tested: a simple Gaussian, the Hulth\'en and the Argonne $v_{18}$ wavefunction.
The Wigner function formalism is applied for the first time to the Hulth\'en and Argonne $v_{18}$ wavefunctions. The Argonne $v_{18}$ wavefunction, which is anchored to a realistic description of the nucleon-nucleon interactions, provides the best agreement with the deuteron spectra measured by ALICE. The predictions obtained with the Argonne $v_{18}$ wavefunction are compared with those obtained with a $\chi$EFT (N$^4$LO) one and they are found to be in excellent agreement. This suggests that the production of the deuteron is not affected by extremely short-range interactions, where the two approaches differ.
The good agreement between data and predictions shows that this model, is able to predict deuteron spectra if the correct proton spectra and source sizes are provided as input. Our work shows how important it is to measure the proton production spectra and the size of the emitting source simultaneously for different collision energies. This would allow for a prediction of the production yields of (anti)deuterons for different energies and multiplicities, providing a reliable estimation of antideuterons produced in the collisions of cosmic rays with the interstellar medium, which constitutes the background for the search for dark-matter annihilation with antinuclei in the final state. 

%\backmatter

% \bmhead{Supplementary information}

% If your article has accompanying supplementary file/s please state so here.

% Authors reporting data from electrophoretic gels and blots should supply the full unprocessed scans for key as part of their Supplementary information. This may be requested by the editorial team/s if it is missing.

% Please refer to Journal-level guidance for any specific requirements.
%\backmatter
\vspace{1cm}
\noindent
\textbf{Acknowledgments.}
%\section*{Acknowledgments} 
We are grateful to Norbert Kaiser for the valuable discussions and his essential help in the steps of calculations.

\section*{Declarations}

This work has received funding from the European Research Council (ERC) under the European Union's Horizon 2020 research and innovation programme (Grant Agreement No 950692).
This work has been supported by the Deutsche Forschungsgemeinschaft through grant SFB 1258 ``Neutrinos and Dark Matter in Astro- and Particle Physics''.
This research was supported by the Munich Institute for Astro- and Particle Physics (MIAPP) of the DFG cluster of excellence ``Origin and Structure of the Universe''.

%Some journals require declarations to be submitted in a standardised format. Please check the Instructions for Authors of the journal to which you are submitting to see if you need to complete this section. If yes, your manuscript must contain the following sections under the heading `Declarations':

%\begin{itemize}
%\item Conflict of interest/Competing interests (check journal-specific guidelines for which heading to use)
%\item Ethics approval
%\item Consent for publication
%\item Availability of data and materials
%\item Authors' contributions
%\end{itemize}

%\noindent
%If any of the sections are not relevant to your manuscript, please include the heading and write `Not applicable' for that section.

%\bigskip

%\bigskip\noindent
%Springer journals and proceedings: \url{https://www.springer.com/gp/editorial-policies}

%\bigskip\noindent
%Nature Portfolio journals: \url{https://www.nature.com/nature-research/editorial-policies}

%\bigskip\noindent
%\textit{Scientific Reports}: \url{https://www.nature.com/srep/journal-policies/editorial-policies}

%\bigskip\noindent
%BMC journals: \url{https://www.biomedcentral.com/getpublished/editorial-policies}
%\end{flushleft}
\appendix
\section{Deuteron Wavefunctions}

%\section{}
\label{app:theory}

\begin{figure}[h]
    \centering
\includegraphics[width=0.45\textwidth]{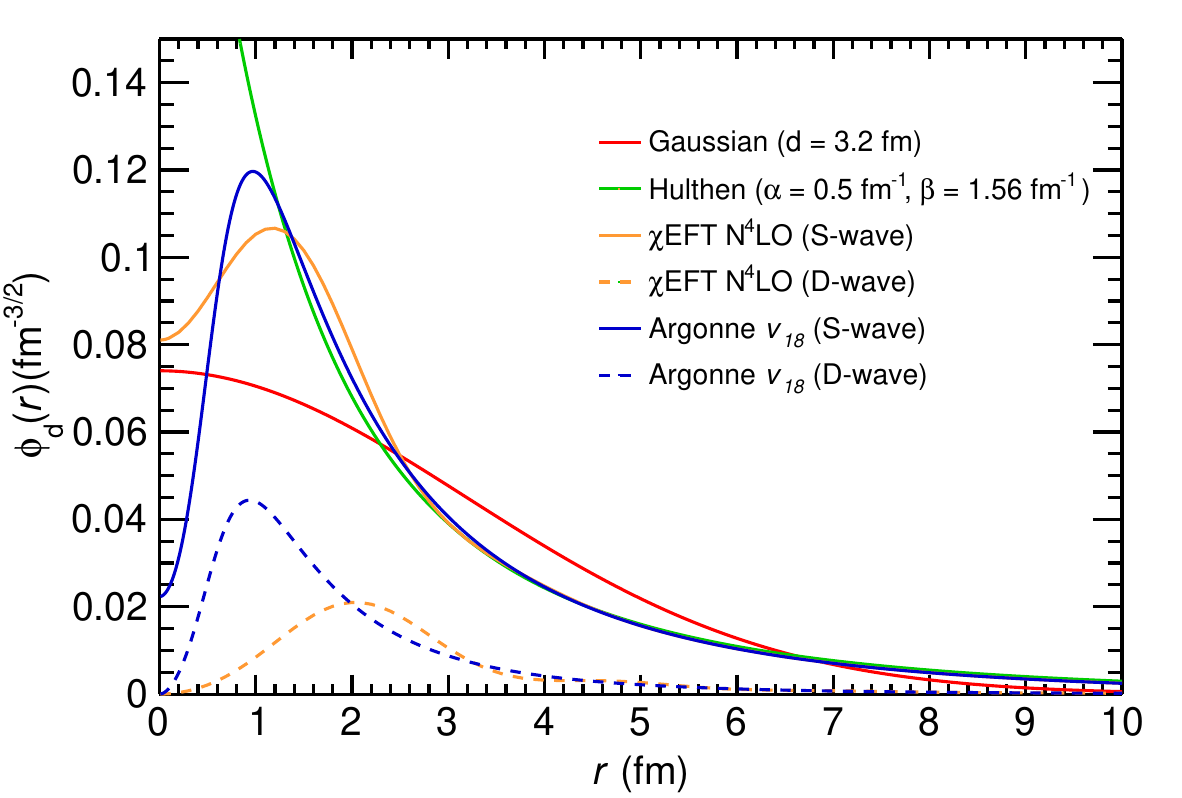}
    \caption{Deuteron wave functions using different potential hypotheses, namely Gaussian (red), Hulth\'en (green)~\cite{Heinz:1999rw}, $\chi$EFT $\text{N}^4\text{LO}$ S-wave and D-wave (solid and dashed orange curves, respectively)~\cite{chiEFT}, Argonne $v_{18}$~\cite{PhysRevC.51.38} S-wave and D-wave (solid and dashed blue curves, respectively).}
    \label{fig:wavefunctions}
\end{figure}

In this Appendix, we report the calculation of Wigner densities $\mathcal{D}(\vec{r}, \vec{q})$ for different hypotheses of the deuteron wavefunction. Namely, we will consider a simple Gaussian, the Hulth\'en , the Argonne $v_{18}$ wavefunctions, and $\chi$EFT as shown in Fig.~\ref{fig:wavefunctions}.  All wavefunctions are normalised according to $\int d^3r\, |\phi(\vec{r})|^2 = 1$, thus the effect on the deuteron yields arises from the different shapes of the wavefunction. 

\subsection{Gaussian wavefunction}
The most simple assumption is a single Gaussian wavefunction

\begin{equation}
  \phi_{\mathrm{d}}(r) = \frac{e^{-\frac{r^2}{2d^2}}}{\left(\pi d^2 \right)^{3/4}},
  \label{eq:gaussian}
\end{equation}
where $d$ is a parameter related to the nucleus radius. For this calculation, $d = 3.2$~fm, as in~\cite{coalescence_correlations}. Using Eq.~\ref{eq:deuteron_wigner}, the corresponding Wigner density is

\begin{equation}
\mathcal{D}(\vec{r}, \vec{q}) = 8\,e^{-\frac{d^4 q^2+r^2}{d^2}} \, .
  \label{eq:Wigsimple_gaussian}
\end{equation}

\subsection{Hulth\'en wavefunction}
The Hulth\'en wavefunction represents a more realistic hypothesis for the deuteron wavefunction and it is based on the Yukawa theory of interaction. The wavefunction has the form
\begin{equation}
  \phi_{\mathrm{d}}(r) = \sqrt{\frac{\alpha \beta (\alpha + \beta)}{2\pi(\alpha - \beta)^2}} \, \frac{e^{-\alpha r} - e^{-\beta r}}{r},
\end{equation}
where $\alpha = 0.23~\text{fm}^{-1}$ and $\beta = 1.61~\text{fm}^{-1}$ are parameters taken from~\cite{Heinz:1999rw}.
For convenience, the Wigner density is calculated starting from the Fourier transform $\psi(\vec{k})$ of the wavefunction
\begin{equation}
  \psi(\vec{k}) = \frac{\sqrt{\alpha+\beta}
  }{\pi (\alpha-\beta)}\left(\frac{1}{k^2+\alpha^2}-\frac{1}{k^2+\beta^2 }\right).
  \label{eq:wigner_fourier}
\end{equation}
In Fourier space, Eq.~\ref{eq:deuteron_wigner} has the form
\begin{multline}
  \mathcal{D}(\vec{r}, \vec{q}) \,=\, \int \mathrm{d}^{3}\xi \, \int \mathrm{d}^{3}k_1 \,\int \mathrm{d}^{3}k_2\,\psi^*(\vec{k}_2)\,\psi(\vec{k}_1) \times \\
    \times \,e^{-i \left[\, \vec{q} \cdot\vec{\xi}+\vec{k}_2\cdot(\vec{r}-\vec{\xi}/2)\,+\,\vec{k}_2\cdot(\vec{r}-\vec{\xi}/2)\right]}.
  \label{eq:Fourier_deuteron_wigner}
\end{multline}
Using the substitutions $\vec{k}_2 = 2\vec{q}+\vec{k}_1$ and $\vec{k}_1 = \vec k+\vec q$, and integrating over $\vec{\xi}$ and $\vec{k}_{2}$, one obtains the following expression
\begin{multline}
  \mathcal{D}(\vec{r}, \vec{q}) \, = \,8 \, \int \mathrm{d}^{3}k \, e^{2 i \vec{r} \cdot\vec{k}} \, \psi^*(\vec{q}-\vec{k}) \, \psi(\vec{q}+\vec{k}) \, =\\
  = \, \int \mathrm{d}^{3}k \, e^{i \vec{r} \cdot\vec{k}} \, \psi^*(\vec{q}-\vec{k}/2) \, \psi(\vec{q}+\vec{k}/2) \, .
\end{multline}
The integral depends on the angle $\theta$ between $\vec{r}$ and $\vec{k}$. To eliminate this dependence on the angle $\theta$, the angular average over $\theta$ is performed using $\sin(\theta)$. With these simplifications, the Wigner density of the Hulth\'en wavefunction becomes
\begin{multline}
  \mathcal{D}(\vec{r}, \vec{q}) \, = \, \frac{4 (\alpha+\beta)^2}{\alpha \beta(\alpha-\beta)\pi q r} \times \\
  \times \, \int_{0}^{\infty} \mathrm{d}k \,  \frac{\alpha^2 \beta^2\sin{(2 k r)}}{\alpha^2 +\beta^2+2(k^2+q^2)} \, \times \\
    \times \, \left\{ \frac{1}{k^2+q^2+\alpha^2} \, \ln \left[\frac{(k+q)^2+\alpha^2}{(k-q)^2+\alpha^2}\right] \right. \, -\\
  - \, \left. \frac{1}{k^2+q^2+\beta^2} \, \ln \left[\frac{(k+q)^2+\beta^2}{(k-q)^2+\beta^2}\right] \right\} \, .
\end{multline}

\subsection{Argonne $v_{18}$ wavefunction}
The Argonne $v_{18}$ potential is a phenomenological potential constrained to proton-neutron scattering measurements~\cite{PhysRevC.51.38}. In such a potential, the deuteron wavefunction has the form 
\begin{equation}
  \phi_{\mathrm{d}}(\vec{r}) = \frac{1}{\sqrt{4\pi} \, r}\left[u(r) \, + \, \frac{1}{\sqrt{8}}\,  w(r) \, S_{12}(\hat{r})\right]\, \chi_{1m}\, , 
  \label{eq:deut_wf_Av18}
\end{equation}
where $S_{12}(\hat{r}) = 3\vec{\sigma}_1\cdot \vec{r}\,\vec{\sigma}_2\cdot \vec{r}-\vec{\sigma}_1\cdot\vec{\sigma}_2$ is the spin tensor, $\chi_{1m}$ is a spinor, and $u(r)$ and $w(r)$ are radial S and D wavefunctions, respectively.
We define $\vec{r}_1$ the coordinates of the proton, $\vec{r}_2$ the coordinates of the neutron, $\vec{r} = \frac{\vec{r}_1-\vec{r}_2}{2}$, and $\vec{R} = \frac{\vec{r}_1+\vec{r}_2}{2}$.
\\The spin averaged density for the deuteron is 
\begin{multline}
  \vert\phi_{\mathrm{d}}(\vec{r})\vert^2\,=\,\frac{1}{3}\sum_{m= 0,\pm 1}\phi_{\mathrm{d}}(\vec{r}_1)^{\dagger} \phi_{\mathrm{d}}(\vec{r}_2)=\\ \,=\, 
  \frac{1}{4\pi\,r_1\,r_2}\Big\{u(r_1)u(r_2)\,+\\+\left.\,w(r_1)w(r_2)\frac{1}{2}\left[3\left(\hat{r}_1\cdot\hat{r}_2\right)^2-1\right] \right\},
  \label{eq:deut_density_Av18}
\end{multline}
and the wavefunction is normalised as
\begin{equation}
 \int \mathrm{d}^3r\, \vert\phi_{\mathrm{d}}(\vec{r})\vert^2 = \int \mathrm{d}^3r \, \frac{1}{4\pi \, r^2}\left[u^2(r) \, + \, w^2(r)\right] = 1 \, .
  \label{eq:deut_density_normalisation_Av18}
\end{equation}
In the previous integral, the contribution of the first addend is dominant, as the first part of the integral is equal to 0.9424~\cite{PhysRevC.51.38}. Since the Argonne $v_{18}$ potential has only a numerical evaluation and no analytical form, an analytical form of its Wigner density is obtained through a fit to the numerical values of $u(r)/r$ and $w(r)/r$. The fit is performed using the function
\begin{equation}
  F(r) = \frac{N_1 a}{\pi(a^2+r^2)}+\frac{N_2b}{\pi(b^2+(r-c)^2)}+N_3 e^{-\frac{r^2}{f}} \, ,
\end{equation}
where $N_1$, $N_2$, $N_3$, $a$, $b$, $c$, and $f$ are fit parameters. $F(r)$ can describe both $u(r)/r$ and $w(r)/r$ individually. Therefore, $u(r)/r$ and $w(r)/r$ are fitted separately and two different sets of fit parameters are obtained for the S and D wave components, respectively. $F(r)$ describes the shape of $u(r)/r$ in the range \mbox{$0 < r < 15$~fm} with  a $\chi^2_{\mathrm{ndf}}\sim 6.83 \cdot 10^{-8}$ and $w(r)/r$ with \mbox{$\chi^2_{\mathrm{ndf}}\sim 1.3 \cdot 10^{-10}$} in the same range. The fit parameters for the S and D waves are reported in Tab.~\ref{table:fitpars}. The corresponding Wigner density for $F(r)$ has the form
\begin{equation}
  \mathcal{D}(\vec{r}, \vec{q}) = \frac{1}{8 \pi}\left(T_1+T_2+T_3+T_4+T_5+T_6\right) \,,
\end{equation}
where the terms  $T_1$, $T_2$, $T_3$, $T_4$, $T_5$, and $T_6$ are defined as
\begin{equation}
    T_1\,=\,(2\pi f)^{3/2} \, N_3^2 \, e^{-\frac{f^2 q^2+4 r^2}{2f}}\,,
\end{equation}
\begin{multline}
    T_2\,=\,\frac{16 a^2 N_1^2 }{\pi q r}\int_{0}^{\infty}\mathrm{d}\xi \, \frac{\sin (q \xi)}{{4 a^2+4 r^2+\xi^2}}\, \times \\
    \, \times \, \ln\left[\frac{4 a^2+(\xi+2 r)^2}{4 a^2+(\xi-2 r)^2}\right]\,,
\end{multline}
\begin{multline}
    T_3\,=\,\frac{8 a N_1 N_3 }{q r}\int_{0}^{\infty}\mathrm{d}\xi\, \sin (q \xi)\,e^{-\frac{2 a^2+4 r^2+\xi^2}{2 f}}\, \times \\
    \times \, \left[\text{Ei}\left(\frac{4 a^2+(2 r+\xi)^2}{4 f}\right) \, - \right. \\
    - \, \left.\text{Ei}\left(\frac{4 a^2+(\xi-2 r)^2}{4 f}\right)\right]\,,
\end{multline}
\begin{multline}
    T_4\,=\,\frac{4 b N_2 N_3}{q}\int_{0}^{\infty}\int_{-1}^{1}\mathrm{d}\xi\,\mathrm{d}\gamma\, \sin (q \xi) \,\xi \, \times \\
    \times \, \left(\frac{e^{-\frac{4 r^2+4 \gamma  r \xi+\xi^2}{4 f}}}{b^2+c^2-c \sqrt{4 r^2-4 \gamma  r \xi+\xi^2}+r^2-\gamma  r \xi+\frac{\xi^2}{4}} \, \right. + \\
    \left. + \, \frac{e^{-\frac{4 r^2-4 \gamma  r \xi+\xi^2}{4 f}}}{b^2+c^2-c \sqrt{4 r^2+4 \gamma  r \xi+\xi^2}+r^2+\gamma  r \xi+\frac{\xi^2}{4}}\right)\,,
\end{multline}
\begin{multline}
    T_5 \,=\,\frac{4 a b N_1 N_2}{\pi  q}\int_{0}^{\infty}\int_{-1}^{1}\mathrm{d}\xi\,\mathrm{d}\gamma\, \sin (q \xi) \,\xi \,\times \\
    \times \,\left[\frac{\left(a^2+r^2+\gamma  r \xi+\frac{\xi^2}{4}\right)^{-1}}{b^2+\left(c-\frac{1}{2} \sqrt{4 r (r-\gamma  \xi)+\xi^2}\right)^2} \, \right. + \\
    \left. + \, \frac{\left(a^2+r^2-\gamma  r \xi+\frac{\xi^2}{4}\right)^{-1}}{b^2+\left(c-\frac{1}{2} \sqrt{4r(r+\gamma  \xi)+\xi^2}\right)^2}\right]\,,
\end{multline}
\begin{multline}
    T_6 \,=\,\frac{4 N_2^2\, b^2}{\pi  q}\int_{0}^{\infty}\int_{-1}^{1}\mathrm{d}\xi\,\mathrm{d}\gamma\, \sin (q \xi) \,\xi \,\times \\
    \times \,\left[\frac{1}{b^2+\left(c-\frac{1}{2} \sqrt{4 r (r+\gamma  \xi)+\xi^2}\right)^2}  \, \right. \times \\
    \left. \times \, \frac{1}{b^2+\left(c-\frac{1}{2} \sqrt{4 r (r-\gamma  \xi)+\xi^2}\right)^2} \right]\,.
\end{multline}
In the previous equations, $\text{Ei}$ is an exponential integral defined as $\text{Ei}(x) = \int_{x}^{\infty} \mathrm{d}t\,e^{-t}/t $.
\begin{center}
\begin{table}[h]
\caption{Fit parameters for $F(r)$ obtained from the numeric values of $u(r)/r$ ($2^{nd}$ column ) and  $w(r)/r$ ($3^{rd}$ column) in the range \mbox{$0 < r < 15$~fm}.}
\centering
\begin{tabular}{|c|c|c|} 
 \hline
Fit parameters & for $u(r)/r$& for $w(r)/r$\\
\hline
$N_1$ & 0.81370516 & -0.34242388\\
$N_2$ & 4.49712863 &1.0973295\\
$N_3$ & -0.68798139 &-0.25201684 \\
$a$ & -10.82747628 & 4.33930564\\ 
$b$ & 1.68243617 &1.28156015\\ 
$c$ & -0.40957858 &0.22952727\\ 
$f$ & 0.39633979 &0.42620769\\ 
\hline
\end{tabular}
\label{table:fitpars}
\end{table}
\end{center}
\subsection{Chiral Effective Field Theory wavefunction}
The Chiral Effective Field Theory (Chiral EFT or $\chi$EFT) is a theoretical framework used to study the low-energy QCD, such as atomic nuclei or hadrons (protons and neutrons). The Chiral EFT approach is systematic in the sense that the various contributions to a particular dynamical process can be arranged as an expansion in terms of a suitable expansion parameter. The expansion parameter is chosen to be the ratio of a typical low momentum (soft scale) to the chiral symmetry breaking scale ($\Lambda_{\chi}\,\sim$ 1 GeV, hard scale). The systematic expansion allows for the derivation of low-energy observables with controlled uncertainties. The deuteron wavefunction is obtained using the $NN$ potentials through five orders of Chiral EFT, ranging from leading order (LO) to next-to-next-to-next-to-next-to-leading order ($\text{N}^4\text{LO}$)~\cite{chiEFT}, with the normalisation defined in~\cite{chiEFTnorm}. A cutoff at $\Lambda_c = 500$~MeV is used. 
As in the case of Argonne $v_{18}$, the wavefunction is composed by two components $u(r)$ and $w(r)$, which correspond to the radial S wave and to the radial D wave, respectively. The two components are shown separately in Fig.~\ref{fig:wavefunctions}. Also in this case, only the numerical values of $u(r)/r$ and $w(r)/r$ are available and an analytic expression of the wavefunction is obtained with a fit, using the function

\begin{multline}
  \mathcal{F}(r) = \frac{N_0}{\left(a^2 r^2-r_0^2\right)^2b^{-1}+c^2 r_0^2} + \\
  + \sum_{i = 1}^{3}\frac{N_i \alpha_i}{\pi((r-\beta_i)^2+\alpha_i^2)} \, ,
\end{multline}
where $N_0$, $N_i$, $\alpha_i$, $\beta_i$, $a$, $b$, $c$, and $r_0$ are fit parameters. For $w(r)/r$, only the second term of $\mathcal{F}(r)$ is used since it is sufficient to describe the numerical values properly. $u(r)/r$ and $w(r)/r$ are fitted separately and the two sets of fit parameters are reported in Tab.~\ref{table:fitparsChiralEFT}. The extracted values of $\chi^2_{\mathrm{ndf}}$ of the fits in the range \mbox{$0 < r < 15$~fm} are $\sim 3.31 \cdot 10^{-8}$ and $\sim 3.23 \cdot 10^{-3}$ for $u(r)/r$ and $w(r)/r$, respectively.\\
The Wigner density of $F(r)$ has the form
\begin{multline}
  \mathcal{D}(\vec{r}, \vec{q}) = \frac{1}{2 q \pi^2}\int_{0}^{\infty}\int_{-1}^{1}\mathrm{d}\zeta\,\mathrm{d}\gamma\, \sin (q \zeta) \,\zeta \times \\ 
  \times\,\left(\kappa_0+\kappa_1+\kappa_2+\kappa_3\right) \,,
\end{multline}
where the terms  $\kappa_0$, $\kappa_1$, $\kappa_2$, and $\kappa_3$ are defined as
%%%% fun begins here!!!!
\begin{multline}
    \kappa_0\,=\frac{4bN_0^2}{\left(4r_0^2-a^2 \left(\zeta ^2+4 r^2+4 \gamma  \zeta  r\right)\right)^2+4 b c^2 r_0^2 }\times \\ 
    \times\frac{1}{\left(4r_0^2-a^2 \left(\zeta ^2+4 r^2-4 \gamma  \zeta  r\right)\right)^2 + 4 bc^2 r_0^2}\,,
\end{multline}
\begin{multline}
 \kappa_1\,=\sum_{i = 1}^{3}\frac{4 N_i^2 \alpha_i^2}{\left(2\beta_i-\sqrt{\zeta ^2+4r^2-4\gamma  \zeta  r}\right)^2+4\alpha_i^2}\times\\\times\frac{1}{\left(2\beta_i-\sqrt{\zeta ^2+4r^2+4\gamma  \zeta  r}\right)^2+4\alpha_i^2}\,,
\end{multline}
\begin{multline}
 \kappa_2 \, = \mathop{\sum_{i=1}^{3} \sum_{j=1}^{3}}_{i\neq j} \, 8 N_i N_j \alpha_i \alpha_j \, \times \\
 \times\left\{\left[\frac{1}{\left(2\beta_i-\sqrt{\zeta ^2+4r^2-4\gamma\zeta r}\right)^2+4\alpha_i^2}\times \right.\right. \\ 
 \left.\left.\times\frac{1}{\left(2\beta_j-\sqrt{\zeta ^2+4r^2+4\gamma \zeta r}\right)^2+4\alpha_j^2}\right]\right. + \\
 + \left.\left[\frac{1}{\left(2\beta_i-\sqrt{\zeta ^2+4r^2+4\gamma \zeta r}\right)^2+4\alpha_i^2}\times\right.\right. \\
 \left.\left.\times\frac{1}{\left(2\beta_j-\sqrt{\zeta^{2} + 4r^{2} - 4 \gamma \zeta r}\right)^2+4\alpha_j^2}\right]\right\}\,,
\end{multline}
\begin{multline}
 \kappa_3\,= \sum_{i=1}^{3} 4\,\pi N_0\,N_i \alpha_i\,\times\\\times\left\{\left[\frac{1}{\left(2\beta_i-\sqrt{\zeta ^2+4r^2-4\gamma\zeta r}\right)^2+4\alpha_i^2}\times \right.\right.\\\left.\left.\times\frac{4b}{\left(4 r_0^2-
 a^2 \left(\zeta ^2+4 r^2+4 \gamma  \zeta  r\right)\right)^2+4 b c^2 r_0^2}\right]\right.+\\+\left.\left[\frac{1}{\left(2\beta_i-\sqrt{\zeta ^2+4r^2+4\gamma\zeta r}\right)^2+4\alpha_i^2}\times \right.\right.\\\left.\left.\times\frac{4b}{\left(4 r_0^2-
 a^2 \left(\zeta ^2+4 r^2-4 \gamma\zeta r\right)\right)^2+4 b c^2 r_0^2}\right]\right\}\,.
\end{multline}
\begin{center}
\begin{table}[h]
\caption{Fit parameters for $\mathcal{F}(r)$ obtained from the numeric values of $u(r)/r$ ($2^{nd}$ column ) and  $w(r)/r$ ($3^{rd}$ column) in the range \mbox{$0 < r < 15$~fm}.}
\centering
\begin{tabular}{|c|c|c|} 
 \hline
Fit parameters & for $u(r)/r$& for $w(r)/r$\\
\hline
$N_0$ & 14.83063014 & -\\
$N_1$ & 0.3644193 &90.06036202\\
$N_2$ & 0.01876164 &0.22901687 \\
$N_3$& 0.58780443& 90.167747\\ 
$a$ & 2.95678555 &-\\ 
$b$ & 7.03082423 &-\\ 
$c$ & 2.85271022 &-\\ 
$r_0$ & 2.65962623 &-\\ 
$\alpha_1$ & 0.86804832 &1.75803721\\ 
$\alpha_2$ & -2.99220936 &2.55621569\\ 
$\alpha_3$ & 2.51249685&-1.7664106\\ 
$\beta_1$ & 1.81024872 &2.07489033\\ 
$\beta_2$ & 12.77230151 &4.11299107\\
$\beta_3$ & 2.95031591&2.0759802\\

\hline
\end{tabular}
\label{table:fitparsChiralEFT}
\end{table}
\end{center}
\section{Angular correlations}
\label{app:AngularCorrelation}
\begin{figure}[!h]
    \centering
    \includegraphics[width=0.49\textwidth]{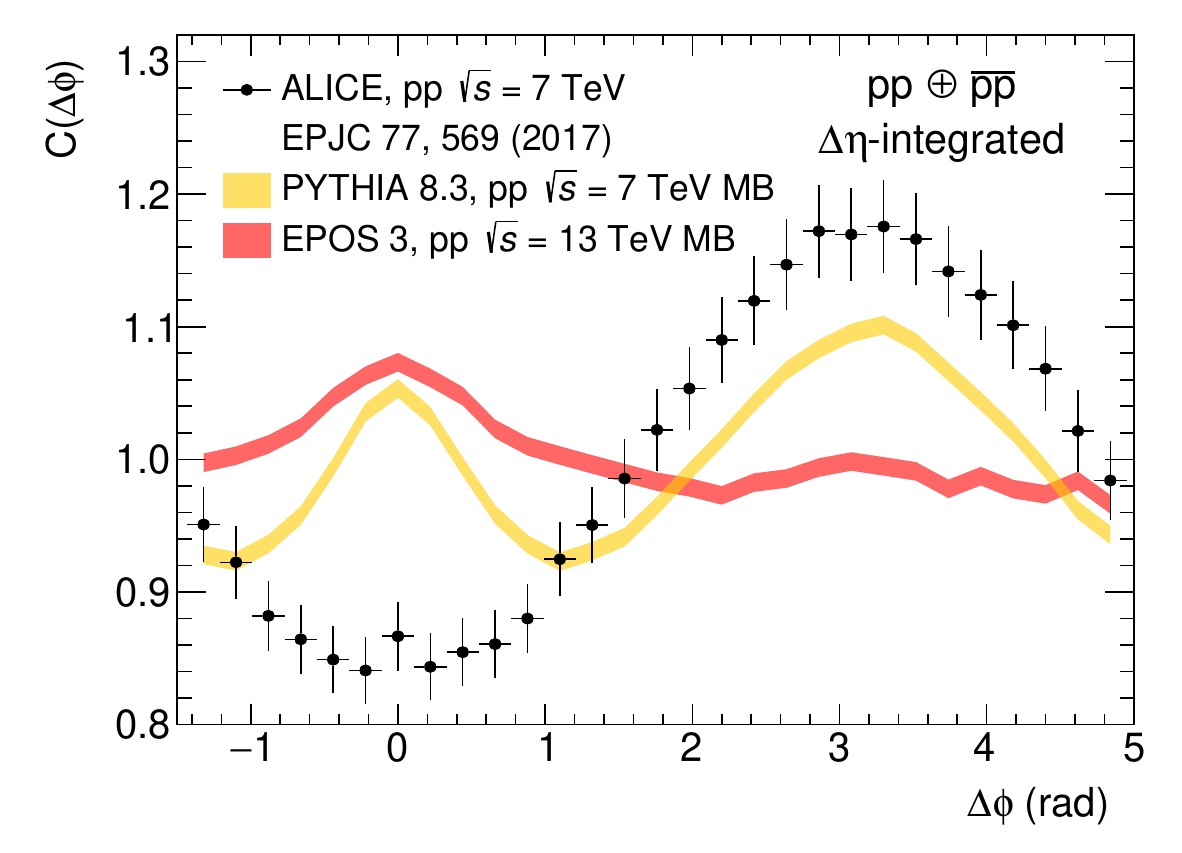}
    \caption{$\Delta\eta$-integrated $\Delta\varphi$ $pp\oplus\Bar{p}\Bar{p}$ correlation function $C(\Delta\varphi)$ of (anti)proton pairs measured by ALICE (black points) and predictions by EPOS 3 (red band) and PYTHIA 8.3 Monash 2013 (yellow band)~\cite{DphiDeta7TeVMB}.}
    \label{fig:DeltaPhi}
\end{figure}

In this Appendix, we discuss the effect of angular correlations, namely $\Delta\varphi\Delta\eta$, on the deuteron predictions. In Fig.~\ref{fig:DeltaPhi} the $\Delta\eta$-integrated $\Delta\varphi$ correlation function $C(\Delta\varphi)$ of sign-like (anti)proton pairs measured by ALICE is shown and compared to predictions by EPOS 3 and PYTHIA 8.3 with the Monash 2013 tune~\cite{DphiDeta7TeVMB}. Note that, while the EPOS 3 prediction was obtained from pp collisions at $\sqrt{s}=$ 13~TeV, the ALICE measurement and PYTHIA prediction were obtained from pp collisions at $\sqrt{s}=$ 7~TeV. However, no qualitative difference between the predictions at these different energies is expected. It is noteworthy that at $\Delta\varphi$ close to zero ALICE measures a depletion in the correlation function, while both models predict a peak. On the contrary, the away-side peak around $\Delta\varphi=\pi$ is much more pronounced in the ALICE measurement compared to Monte Carlo studies. These discrepancies show a fundamental flaw in the production mechanism of baryons in these models. This means that it is impossible to properly correct these effects a posteriori. Instead, a rework of the hadron production mechanism inside the event generators would be required. A worst-case estimation of the effect on deuteron spectra can be performed by reweighting the p-n pairs according to the ratio of measured and predicted $C(\Delta\varphi)$. Such a conservative estimate would lower the overall deuteron yield by no more than 20\%. This effect is neglected in this study.

\section{correlation of q and r}
\label{app:QRCorrelation}
\begin{figure}[!hbt]
    \centering    \includegraphics[width=0.55\textwidth]{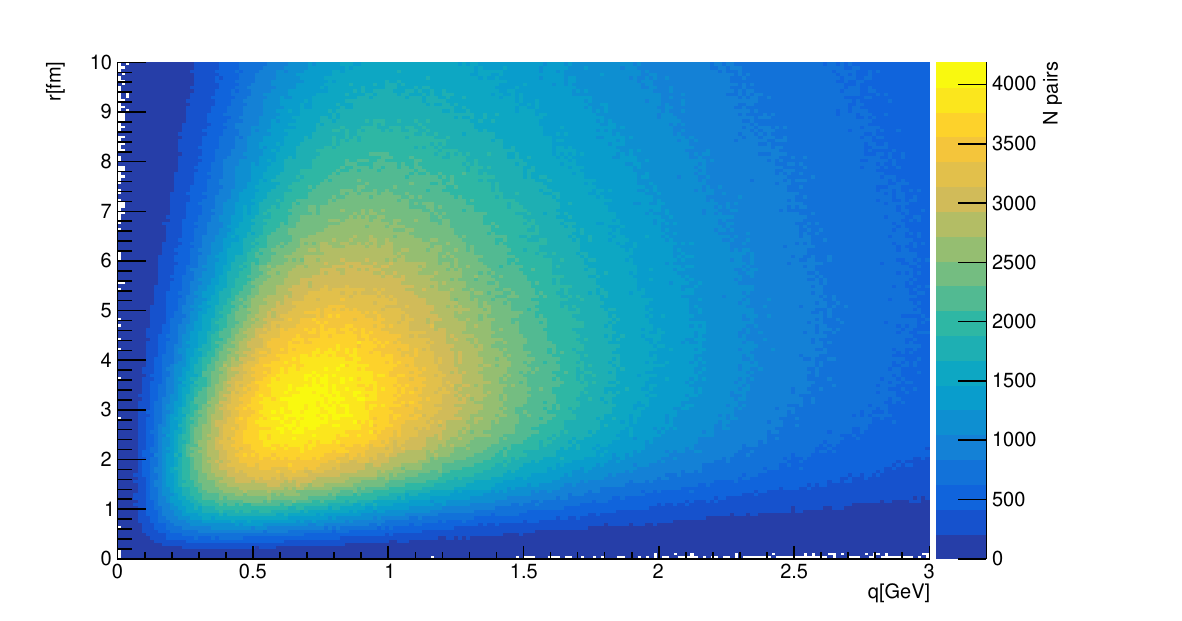}
    \caption{Distribution of relative momenta and distances of proton-neutron pairs in EPOS 3.}
    \label{fig:QRCorrelation}
\end{figure}

Fig.~\ref{fig:QRCorrelation} shows the distribution of relative momentum \textit{q} and distance \textit{r} for proton-neutron pairs, evaluated in the pair rest frame. Clearly visible is a positive correlations between \textit{q} and \textit{r}, where small relative momenta are preferred for pairs with small distances. This effect enhances the deuteron yield by roughly 10\% since the phase-space region interesting for coalescence (\textit{q} $\lesssim 0.5$ GeV/c, \textit{r} $\lesssim 2$ fm) is more populated than a sample with no correlation.

\bibliographystyle{ieeetr}
\bibliography{References}

\end{document}